\documentclass[a4paper,12pt,preprint]{elsarticle}

\usepackage[T1]{fontenc}

\usepackage{amsmath,amssymb}
\usepackage{graphicx}
\usepackage[dvipsnames, svgnames, table]{xcolor}

\usepackage{siunitx}
\usepackage{setspace}

\usepackage{textcomp}

\usepackage[english]{babel}

\usepackage[babel]{csquotes}

\usepackage{float}

\usepackage{subcaption}
\usepackage[inline]{enumitem}
\setlist{nolistsep}

\usepackage{hyperref}
\hypersetup{
    breaklinks = true,
    colorlinks = true,
    bookmarksopen = true,
}
\usepackage[section, above, below]{placeins}

\usepackage[noabbrev, nameinlink]{cleveref}

\def\be{\begin{equation}}
\def\ee{\end{equation}}
\def\ba{\begin{array}}
\def\ea{\end{array}}

\journal{Composites Part B: Engineering}

\title{Periodicity and Image Registration for Yarn Path Extraction in Large 3D Textiles}

\author[1,2,3]{Hafsa El Herichi}
\author[1,3]{Arturo Mendoza \corref{cor1}}
\author[4]{Yanneck Wielhorski}
\author[2]{Hugues Talbot}
\author[1]{Stéphane Roux}
\cortext[cor1]{Corresponding author.}
\ead{arturo.mendoza-quispe@safrangroup.com}

\address[1]{Universit\'e Paris-Saclay, CentraleSup\'elec, ENS Paris-Saclay, CNRS, LMPS, \\4, Avenue des Sciences, 91192 Gif-sur-Yvette, France}
\address[2]{Universit\'e Paris-Saclay, CentraleSup\'elec, Inria, CVN, \\9 Rue Joliot Curie, 91190 Gif-sur-Yvette, France}
\address[3]{Safran Tech, Rue des Jeunes Bois, 78772 Magny-les-Hameaux, France}
\address[4]{Safran Aircraft Engines, Rond-point R\'en\'e Ravaud - R\'eau, 77550 Moissy-Cramayel, France}

\begin{document}

\begin{frontmatter}

\begin{abstract}
Accurate identification of yarns in X-ray computed tomography volumes remains a critical and complex step in generating high-fidelity numerical models of woven composites.  This work introduces a tracking framework that leverages the intrinsic periodicity of woven architectures to transform a complex, large-scale segmentation problem into the annotation of a single representative unit cell.

The approach first exploits the periodic nature of the weave to extract a representative unit cell from the volumetric data and segment it to provide a reference description of the yarn geometry.
Digital Volume Correlation (DVC) is then performed between an idealised periodic volume obtained by replication of the unit cell and the real composite volume, yielding a three-dimensional displacement field that captures geometric deviations from ideal periodicity.
The unit-cell yarn segmentation is propagated to the full volume by exploiting the periodicity of the reference volume, and the annotations are transported to the actual volume using the DVC-derived displacement field.

Results obtained on real composite datasets demonstrate the ability of the method to accurately recover warp and weft yarn architectures with minimal input, opening new perspectives for efficient and scalable textile composite characterisation.
\end{abstract}

\end{frontmatter}

\section{Introduction}

Composite materials have revolutionised modern engineering by offering superior specific mechanical properties compared to traditional materials~\cite{LIU2023111176}.
Their high strength-to-weight ratio, corrosion resistance, and design flexibility have led to widespread adoption in aerospace, automotive, and civil engineering applications~\cite{Khan_automobile,Khan_engineering}.
In aero-engines, the integration of 3D woven carbon fibre reinforced polymers (CFRPs) has enabled substantial performance gains, notably in the LEAP engine, where composite fan blades and fan case contribute to significant fuel savings and weight reduction~\cite{ZHANG2023110463}.

The geometry of such blades reflects strongly heterogeneous mechanical requirements.
While the airfoil region must remain thin to ensure aerodynamic efficiency, the root is considerably thicker and incorporates a denser woven architecture~\cite{patent_safran}.
This region is critical for resisting the high centrifugal forces generated during rotation, which necessitates enhanced stiffness, strength, fatigue resistance, and the ability to transfer loads to the disk.
As a consequence of the increased mechanical resistance, this local densification increases the complexity of X-ray computed tomography image analysis.
In these dense regions, yarns are tightly packed, and image contrast is reduced even for dry preforms, making the reliable identification and tracking of individual warp and weft yarns particularly challenging, even for experienced annotators.

X-ray CT has become a key tool for reconstructing the internal textile architecture of composites~\cite{NDT_sota,IRANSHAHI2025200516}.
Beyond visualisation, accurate yarn segmentation is essential for generating geometrically faithful mesoscopic models that support multi-scale simulations and predictive mechanical analyses to assess the structural integrity of the parts~\cite{Wang31122022,Chen2023}.
As such, ensuring geometric periodicity is particularly important when constructing representative volume elements intended for homogenisation procedures or simulations involving periodic boundary conditions. Considerable effort is often devoted to recovering or enforcing periodicity from CT-derived geometries before numerical analysis~\cite{sun2025mesoscopic,sun2026direct}, while alternative approaches rely on formulations that weakly enforce periodicity constraints~\cite{guo2025deep}.

Moreover, most existing studies focus on academic-scale coupons acquired at high spatial resolution ~\cite{CAO2025113294,Blusseau_CompPartB_2022,Zheng_CST_2025}, where yarns are well-separated, and contrast conditions are favourable.
In contrast, dense industrial regions at the component scale~\cite{Sinchuk_CompPartA_2024} remain largely unexplored, and robust segmentation strategies capable of handling such configurations are still lacking.

Indeed, a related recent study~\cite{hafsa2026}, conducted on the same experimental dataset as the one considered here, addressed less densely packed regions where individual yarns remain sufficiently distinguishable to enable local tracking strategies.
The present work instead focuses on highly compacted regions in which such yarn-wise tracking becomes impractical, motivating a more global description at the scale of the periodic unit cell.

Digital Volume Correlation (DVC)~\cite{Bay_1999,Buljac_2018} enables the estimation of full-field displacement fields by registering volumetric images and is widely used to measure internal displacement and strain fields in materials~\cite{Roux_2008, Rethore_2012}.

Beyond this conventional framework, several studies have explored correlation-based approaches as tools for geometric comparison and representation transfer in textile composites~\cite{MENDOZA2019735,Mendoza_CS_2019_correlation}, or for aligning real tomographic data with idealised textile architectures~\cite{rubino2024alignment}. Similarly, DVC has also been used as a practical tool for aligning CAD geometries or finite element meshes with reference tomographic volumes~\cite{gras2015identification,fragnaud2024model}.

Rather than being used for strain measurement, DVC is employed here as a geometrically consistent mapping and transport operator between an ideal periodic textile and the real distorted architecture. The proposed approach leverages the approximate periodicity of woven composites, which remains preserved even in dense regions. A representative periodic unit cell is first extracted from the tomographic data and segmented, after which an ideal periodic volume is reconstructed by replicating this unit cell and extending the annotations to an arbitrary size. DVC is then performed between the ideal periodic and real volumes, and the resulting displacement field is used to transfer the yarn annotations onto the real architecture.

By combining periodicity-driven modelling with full-field kinematic mapping, the proposed framework reduces a large-scale segmentation problem to the annotation of a single representative cell, enabling scalable and structurally consistent reconstruction of dense 3D woven composite architectures.

\section{Dataset}

This study focuses on the \enquote{as-woven} textile reinforcement of a LEAP fan blade, prior to forming and matrix impregnation.
X-ray CT imaging at 140~{\textmu}m voxel size is used to observe the part, but due to its considerable size, it was scanned into four distinct lengthwise sections.

Our analysis focuses on the first scan, which captures the root of the blade.
This region is of particular interest because it plays a key structural role, connecting the blade to the hub and supporting important mechanical loads.
From a weaving perspective, it also features a complex arrangement of yarns, with curved paths and a tightly packed yarn structure.
These features make it a challenging but representative area to test and validate our approach.
The scanned volume measures ${796 \times 1848 \times 2383}$~voxels (or ${111 \times 259 \times 335}$~mm$^3$).

\section{Studying the weave periodicity}

\begin{figure}[H]
    \centering
    \begin{subfigure}{0.45\textwidth}
        \centering
       \includegraphics[width=0.9\textwidth]{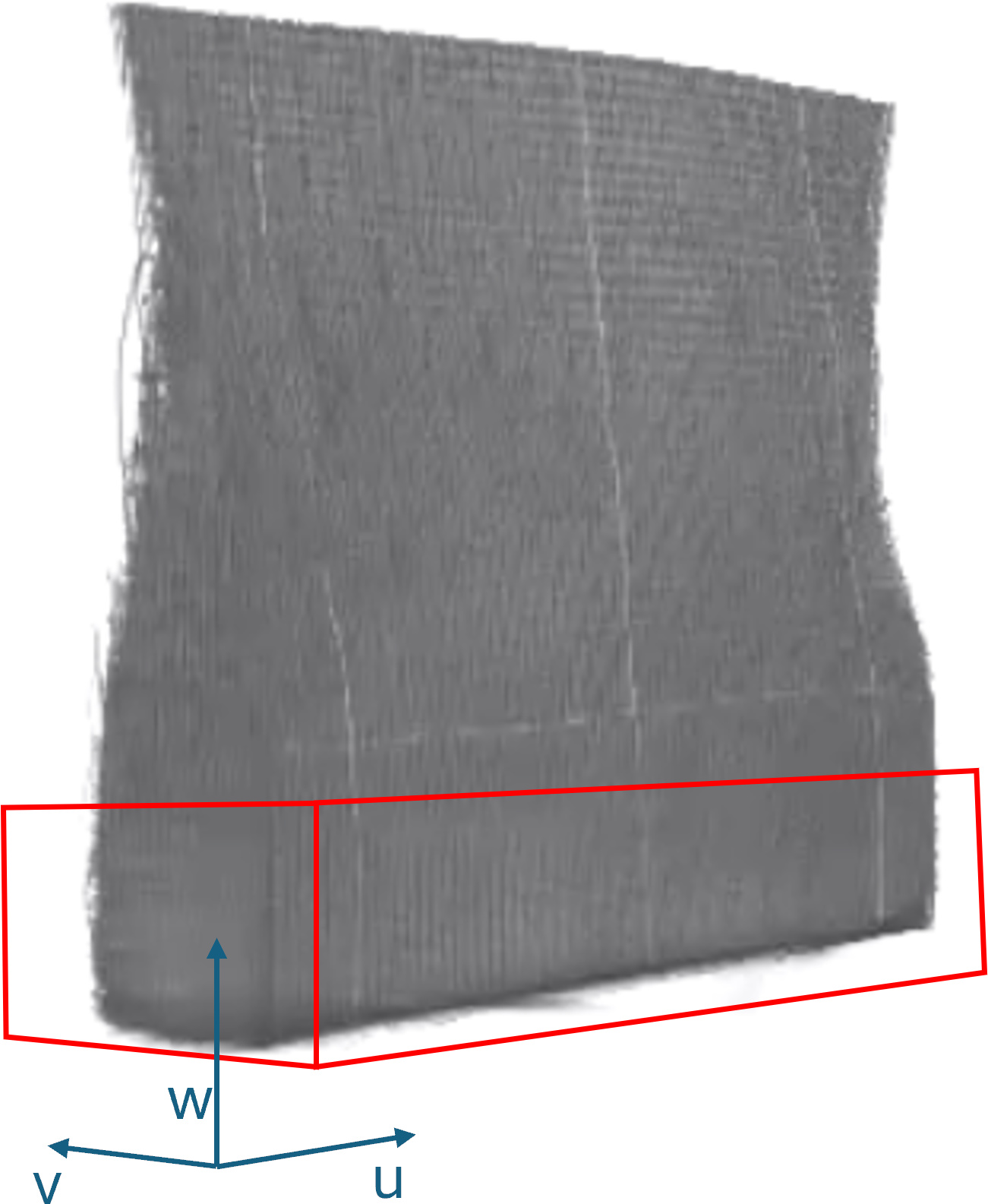}
        \caption{Region of interest.}
        \label{fig:roi_dvc}
    \end{subfigure}
    \hfill
    \begin{subfigure}{0.45\textwidth}
        \centering
        \includegraphics[width=1\textwidth]{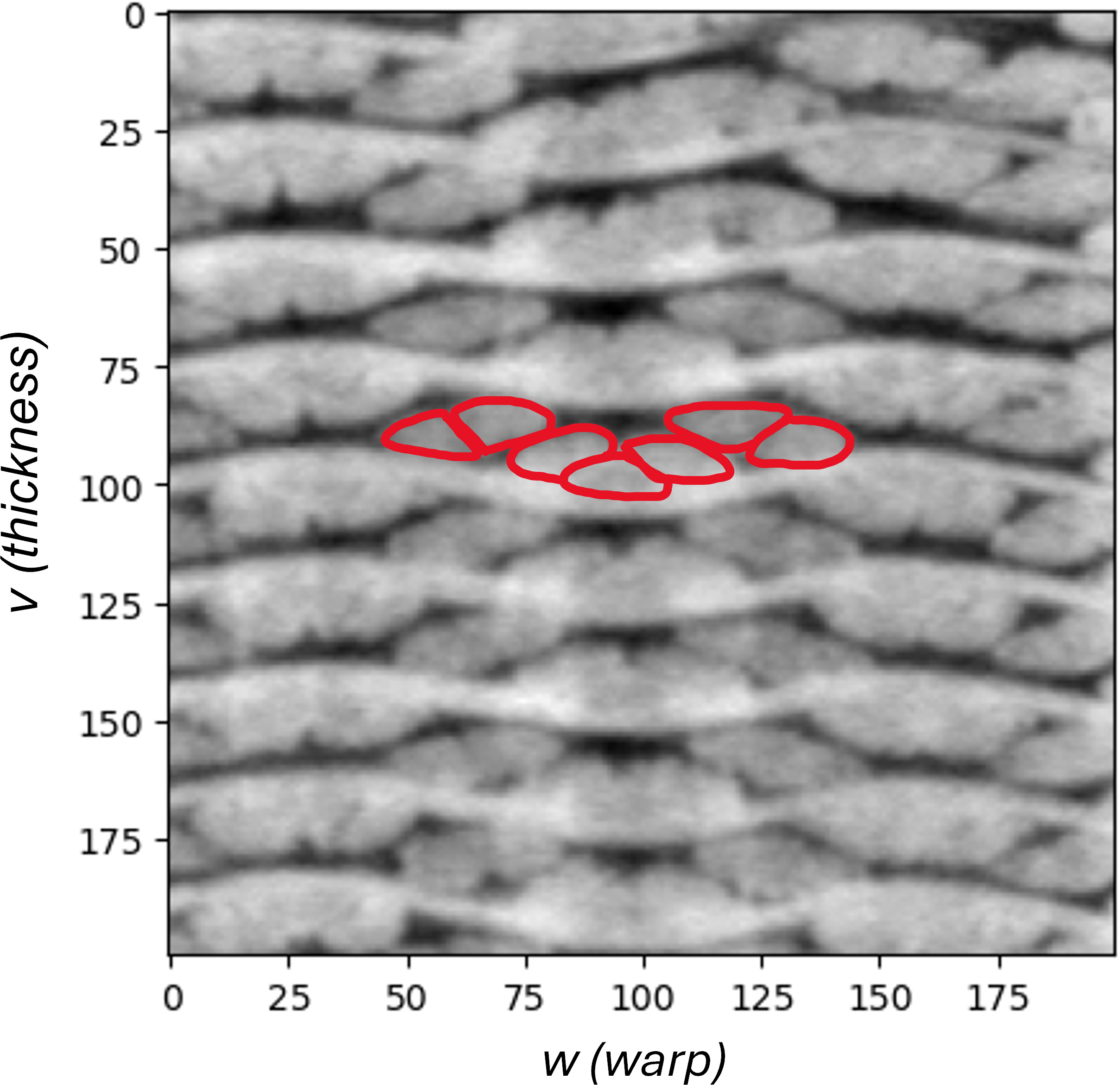}
        \caption{Manually annotated weft yarn cross-sections.}
        \label{fig:weft_yarns_roi}
    \end{subfigure}
    \caption{Region of interest selected for weft yarn analysis and corresponding cross-sections. Weft yarns are aligned along the $u$-axis on average, and warp yarns along the $w$-axis.}
    \label{fig:roi_and_weft}
\end{figure}

The area of focus for this analysis is located at the root of the blade (\Cref{fig:roi_dvc}).
The woven architecture in the analysed region
contains weft yarns of higher tow size (larger number of elementary carbon fibres) than in the upper blade section.
Moreover, the transition from these larger yarns to smaller ones occurs on the boundary of the studied region (towards the blade).

The high number of carbon fibres in these yarns increases segmentation complexity, as their larger size leads to tight packing and unclear cross-sections and boundaries.

In~\Cref{fig:weft_yarns_roi}, manually annotated yarn cross-sections are highlighted for illustration purposes. The yarns are tightly packed, exhibiting varying cross-sectional shapes, and presenting very limited contrast, making the distinction between neighbouring yarns particularly difficult. Due to the complexity of manual annotation under such conditions, this region has remained largely unexplored.

\subsection{Characterisation of the periodic structure}

\begin{figure}[H]
    \centering
    \includegraphics[width=0.48\textwidth]{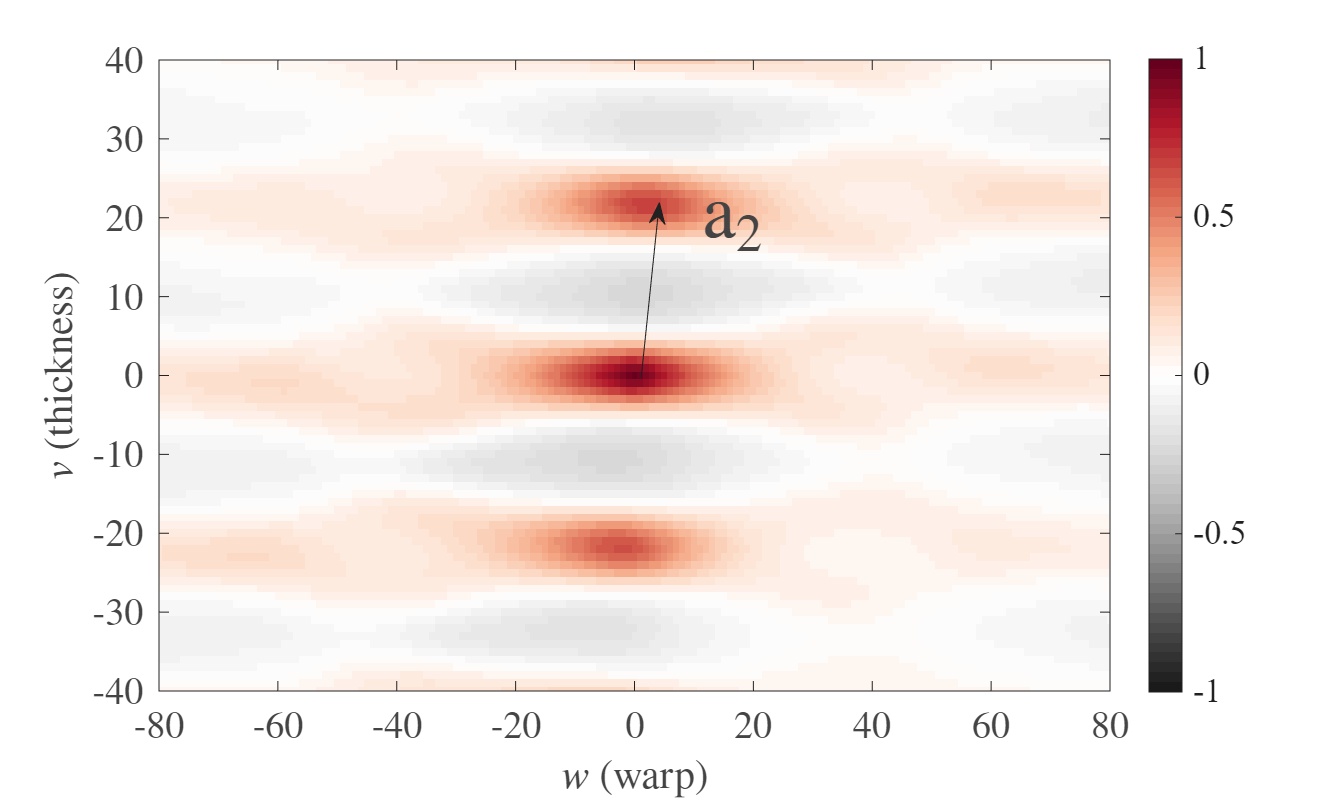}
    \includegraphics[width=0.48\textwidth]{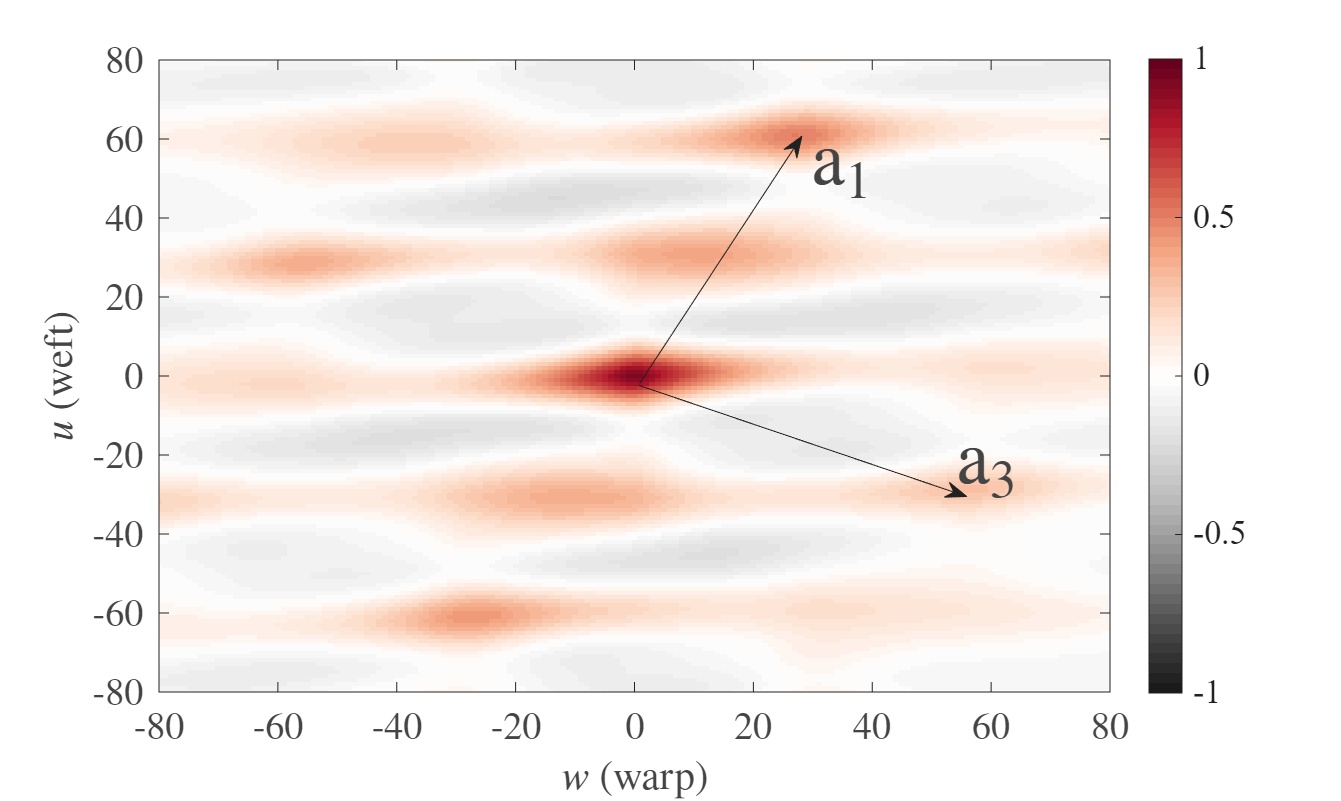}
    \includegraphics[width=0.48\textwidth]{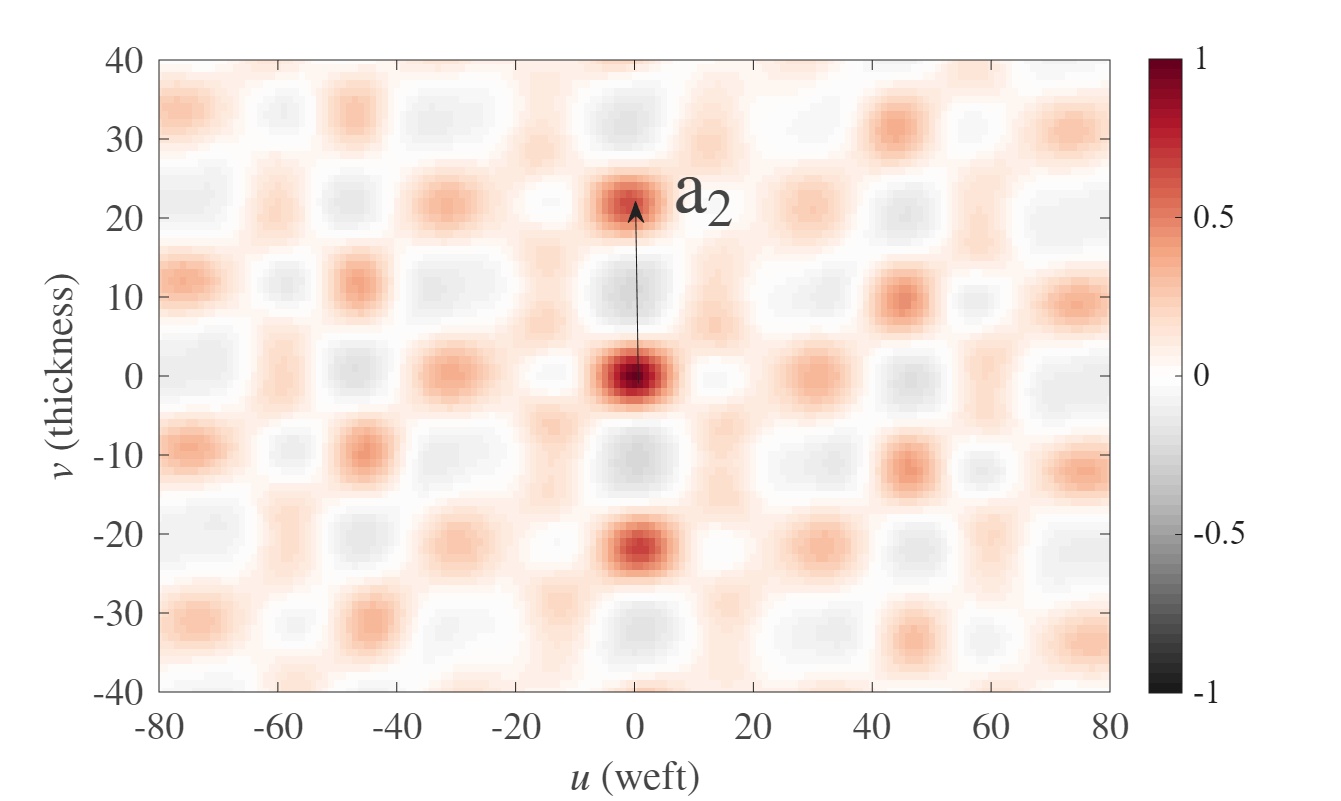}
    \caption{The 3D autocorrelation function is shown along three orthogonal planes $u=0$, $v=0$ and $w=0$.  The three maxima closest to the origin identify the three basic vectors of the periodic cell.    }
    \label{fig:autocorr_peaks}
\end{figure}

The periodic architecture of the weave is characterised using the (three-dimensional) normalised autocorrelation function applied to the region of interest (ROI).
Local maxima of the autocorrelation function correspond to translation vectors that preserve structural similarity, thereby revealing the 3D periodicity of the weave structure. The three main autocorrelation peaks closest to the origin are extracted and shown in \Cref{fig:autocorr_peaks}. This representation reveals the geometric distribution of periodicity vectors within the structure.

These three dominant linearly independent vectors are selected to define the periodic lattice of the weave structure
\be\label{eq:period}\left\{\ba{rcl}
\mathbf{a}_1 &=& ( 61,  0, 26), \\
\mathbf{a}_2 &=& ( -1, 23,  1), \\
\mathbf{a}_3 &=& (-29,  2, 51).
\ea\right.\ee
They form a non-orthogonal basis that spans the fundamental repeating unit.
While these periodicity vectors are not perfectly aligned with the reference image coordinate system $(u,v,w)$, $\mathbf{a}_2$ closely maps to the thickness direction $v$.

\subsection{Constructing a periodic cell and ideal volume}
\label{sec:ideal_volume}

The non-orthogonal $\{\mathbf{a}_1, \mathbf{a}_2, \mathbf{a}_3\}$ basis can be used to construct the linear transformation matrix $\mathbf{T} \in \mathbb{R}^{3 \times 3}$:
\begin{equation}
\mathbf{T} =
\begin{bmatrix}
\mathbf{a}_1 \\ \mathbf{a}_2 \\ \mathbf{a}_3
\end{bmatrix} =
\begin{bmatrix}
a_{1,1} & a_{1,2} & a_{1,3} \\
a_{2,1} & a_{2,2} & a_{2,3} \\
a_{3,1} & a_{3,2} & a_{3,3}
\end{bmatrix},
\end{equation}
which defines a linear mapping between a periodic orthonormal coordinate system $(\mathbf{b}_1, \mathbf{b}_2, \mathbf{b}_3)$ and the reference image coordinate system $(u,v,w)$ such that $\mathbf{a} = \mathbf{T}\mathbf{b}$, where $\mathbf{b} = (\mathbf{b}_1, \mathbf{b}_2, \mathbf{b}_3)^\top$ and $\mathbf{a} = (\mathbf{a}_1, \mathbf{a}_2, \mathbf{a}_3)^\top$.
In this new coordinate system, the periodic unit cell corresponds to $[0,1]^3$ in $(\mathbf{b}_1, \mathbf{b}_2, \mathbf{b}_3)$ coordinates.

The transformation $\mathbf{T}$ is applied to the tomographic region of interest (using interpolation) to produce a new \enquote{distorted} image.
A single unit cell is then extracted from the centre of the tomographic volume to minimise boundary effects.

Due to the approximate periodicity of the textile architecture, neighbouring unit cells are expected to exhibit only limited variations, such that selecting another nearby cell would have little impact on the reconstructed architecture. The choice of a central region is primarily motivated by the fact that periodicity progressively deteriorates near the boundaries of the component and within transition regions between different textile architectures, such as the transition from the dovetail toward the blade. In addition, imaging artefacts, such as beam-hardening effects near specimen boundaries, may further affect the quality of the reconstruction --- and in turn registration --- in these regions.

\begin{figure}[H]
    \centering
    \begin{subfigure}[c]{0.48\textwidth}
        \centering
        \includegraphics[width=\linewidth]        {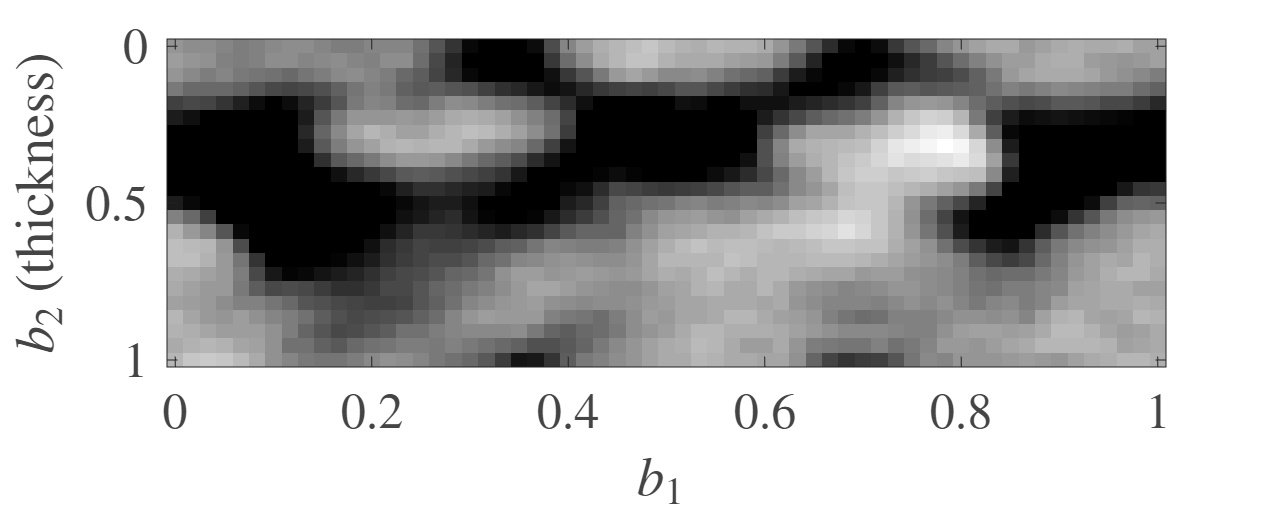}
        \caption{}
    \end{subfigure}
    \hfill
    \begin{subfigure}[c]{0.48\textwidth}
        \centering
        \includegraphics[width=0.9\linewidth]{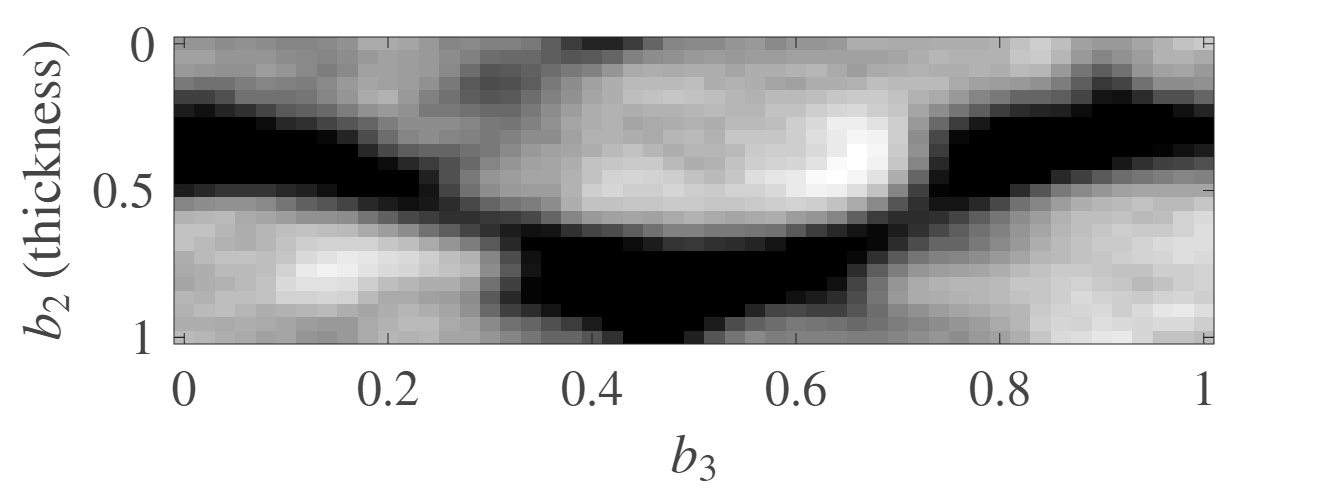}
        \caption{}
    \end{subfigure}
    \hfill
    \begin{subfigure}[c]{0.48\textwidth}
        \centering
        \includegraphics[width=0.8\linewidth]{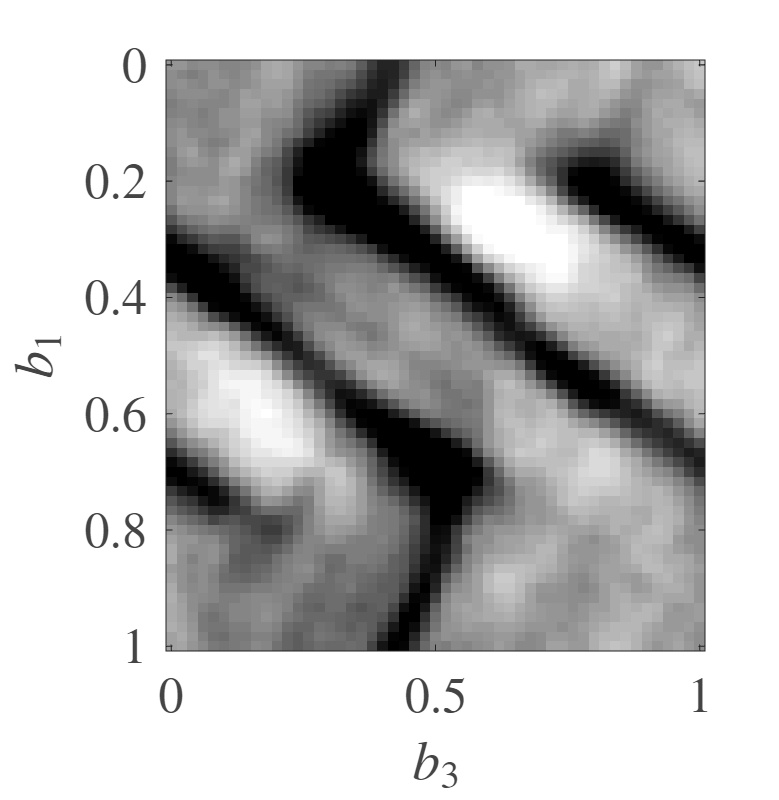}
        \caption{}
    \end{subfigure}
    \caption{Periodic cell extracted from the volume: (a) Normal to the $\mathbf{b}_3$ direction; (b) Normal to the $\mathbf{b}_1$ direction; (c) Normal to the  $\mathbf{b}_2$ direction.
    }
    \label{fig:periodic_cell}
\end{figure}

Two-dimensional slices of this unit cell are shown in~\Cref{fig:periodic_cell}, revealing the internal organisation of warp and weft yarns and highlighting the difficulty of segmenting yarn cross-sections, particularly for weft yarns.

The periodic unit cell constructed above represents the minimal repeating volume element of the woven structure.
A periodic volume is then generated by tiling this unit cell through translations along the image directions $(\mathbf{b}_1, \mathbf{b}_2, \mathbf{b}_3)$. This operation produces a volume exhibiting exact translational periodicity.

This tiled volume can then be interpolated back to the original reference frame $(u,v,w)$ using the inverse linear transformation defined by $\mathbf{T}^{-1}$.
This procedure preserves the intrinsic periodicity of the structure while producing a synthetic volume that is geometrically aligned with the reference image coordinate system.

\section{Segmentation of the unit cell}

\subsection{Warp yarns}

\begin{figure}[H]
    \centering
    \includegraphics[width=0.5\textwidth]{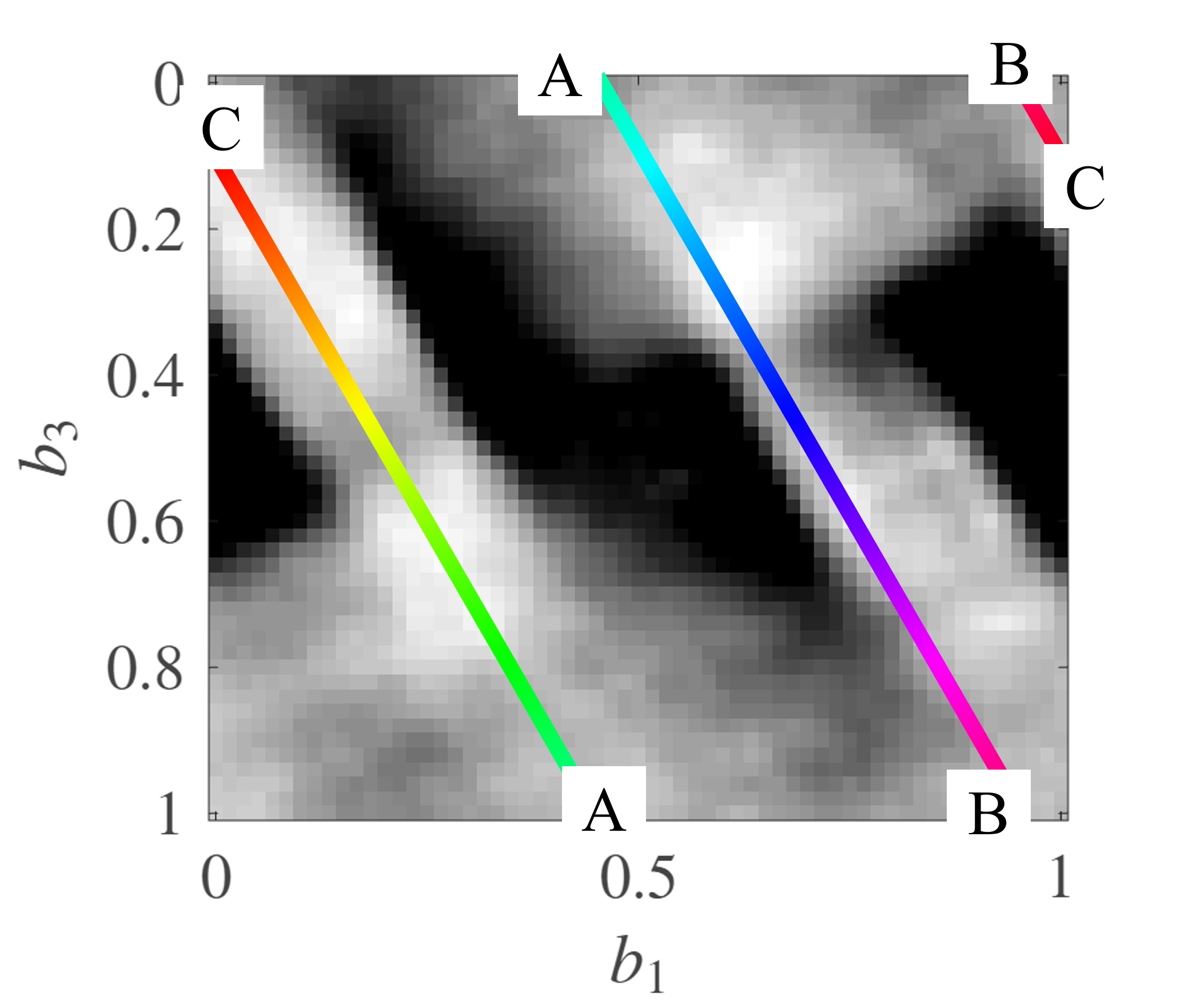}
    \caption{Pattern of a warp yarn on the periodic cell.}
    \label{fig:yarn_periodicity}
\end{figure}

The segmentation of warp yarns within the periodic cell exploits the inherent periodicity of the weave structure.
Although the yarn segments observed across the unit cell are physically distinct, the periodic boundary conditions impose that a yarn exiting one face of the cell must match a yarn entering the opposite face at the exact same position.
In this sense, physically separate segments can be treated as continuous yarns wrapping through the volume.
Therefore, the annotation strategy consists of identifying and grouping these segments as a unified entity.
As illustrated in~\Cref{fig:yarn_periodicity}, a yarn segment traversing the cell from A to B continues in the adjacent cell from B to C, and wraps from C to A.
This continuity across neighbouring cells provides a practical rule for segmentation: segments that are spatially connected through the periodic boundary can be grouped and annotated as a single entity.

The annotations were generated using the detection method introduced in~\cite{hafsa2026}.
This approach first constructs a fundamental mode representing the average cross-sectional shape of the warp and weft yarns observed within the unit cell. These modes are defined using parametric elliptical functions with appropriate geometrical dimensions. A cross-correlation is then computed between the corresponding mode and the tomographic slices in order to determine the in-plane position of the yarns. Starting from an initial slice, tracking from one slice to the next is achieved by identifying the local maximum of the cross-correlation within a restricted neighbourhood, under the assumption that yarn displacements remain limited between consecutive planes. The procedure is applied independently for each yarn orientation (warp and weft) and for each yarn present in the unit cell.

\begin{figure}[H]
    \centering
    \includegraphics[width=\textwidth]{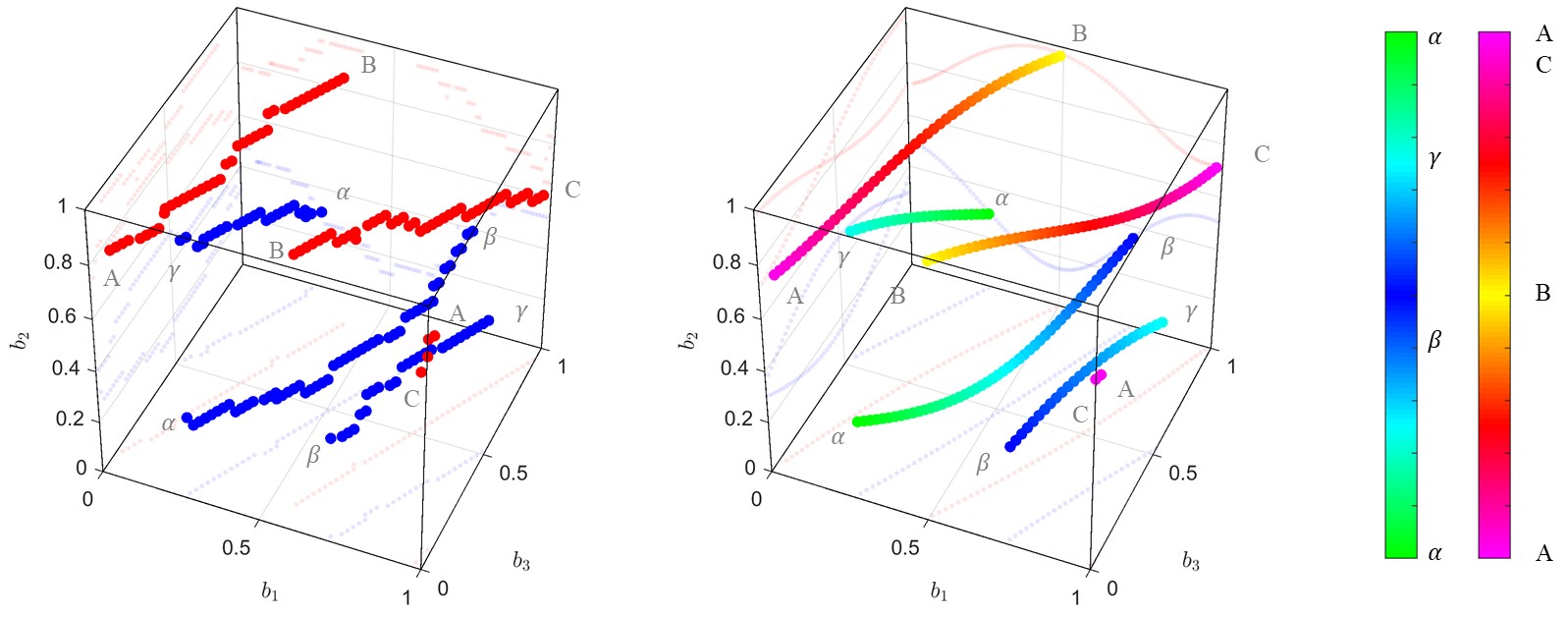}
    \caption{Left: tracked yarn trajectories of two warp yarns in the unit cell; their projections onto the planes are shown to illustrate the periodicity.  Right: fitted sinusoidal trajectories for each yarn.  In both cases, the A-B-C labels indicate the wrap sections for the first yarn, while the $\alpha$-$\beta$-$\gamma$ do so for the second yarn.}
    \label{fig:sinus_curves}
\end{figure}

\Cref{fig:sinus_curves} shows the tracked yarn trajectories. Once the segments belonging to each yarn are identified, a sinusoidal behaviour is observed along their paths. This behaviour becomes apparent when the required jumps between opposite faces of the periodic cell are taken into account. For both yarns, two positive jumps along $b_3$ and one negative jump along $b_1$ are needed to recover a continuous path.

A sinusoidal function with the appropriate periodicity is then fitted to each annotated curve in three dimensions, yielding a smooth and explicitly periodic description of the yarn paths for subsequent propagation steps. Although this fitting ignores small local geometric variations, the proposed framework does not fundamentally rely on sinusoidal parameterisations, and raw tracked paths could also be used provided periodic continuity is ensured. In the present context, where the objective is the large-scale reconstruction of thousands of yarn trajectories for subsequent numerical analyses, a smooth and geometrically consistent representation is preferable.

The trajectories are then generated for the ideal periodic volume by applying the same coordinate mapping procedure described previously in~\Cref{sec:ideal_volume}, so that the warp yarn paths re-align with the original $w$-axis.

\subsection{Weft yarns}

For the segmentation of weft yarns, an adapted strategy is employed because of their denser packing.  This higher density makes it harder to locate the exact boundary between neighbouring yarns, though similar segmentation approaches can still be applied.  Even visual identification of weft yarn boundaries is challenging in the tomographic images.  To address this difficulty, the previously segmented warp yarns information is leveraged as a masking tool.

\begin{figure}[H]
\begin{center}
    \begin{subfigure}{0.6\textwidth}
        \centering
        \includegraphics[width=1\linewidth]{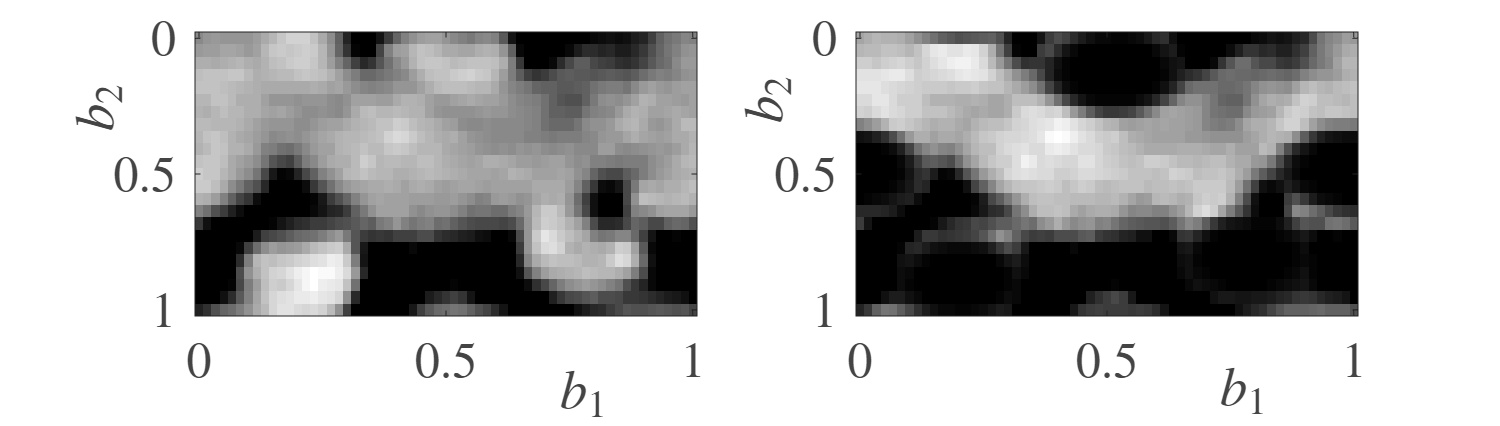}
        \caption{}
    \end{subfigure}
    \hfill
    \begin{subfigure}{0.6\textwidth}
        \centering
        \includegraphics[width=\linewidth]{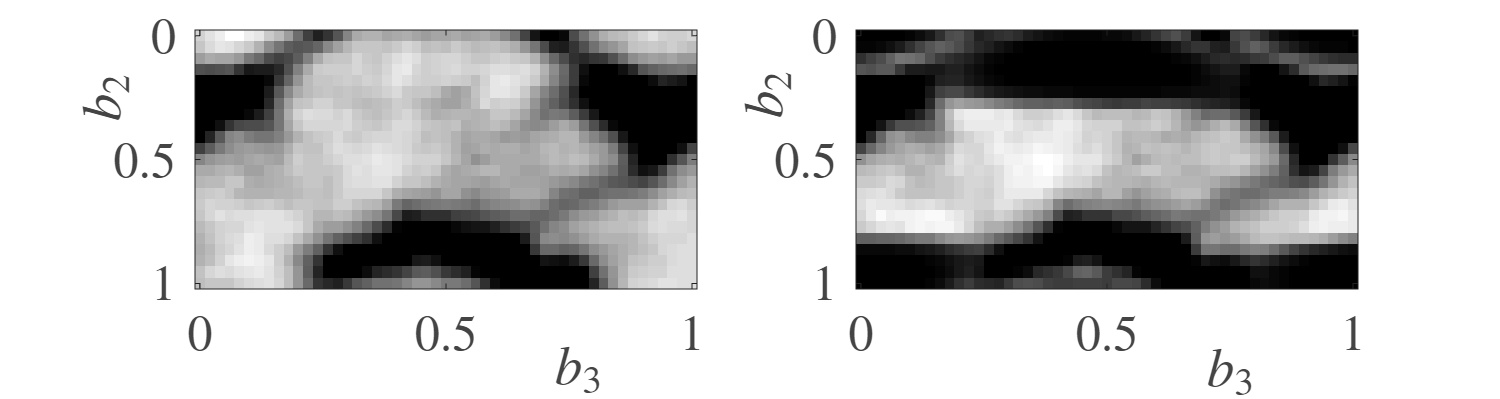}
        \caption{}
    \end{subfigure}
    \hfill
    \begin{subfigure}{0.6\textwidth}
        \centering
        \includegraphics[width=0.8\linewidth]{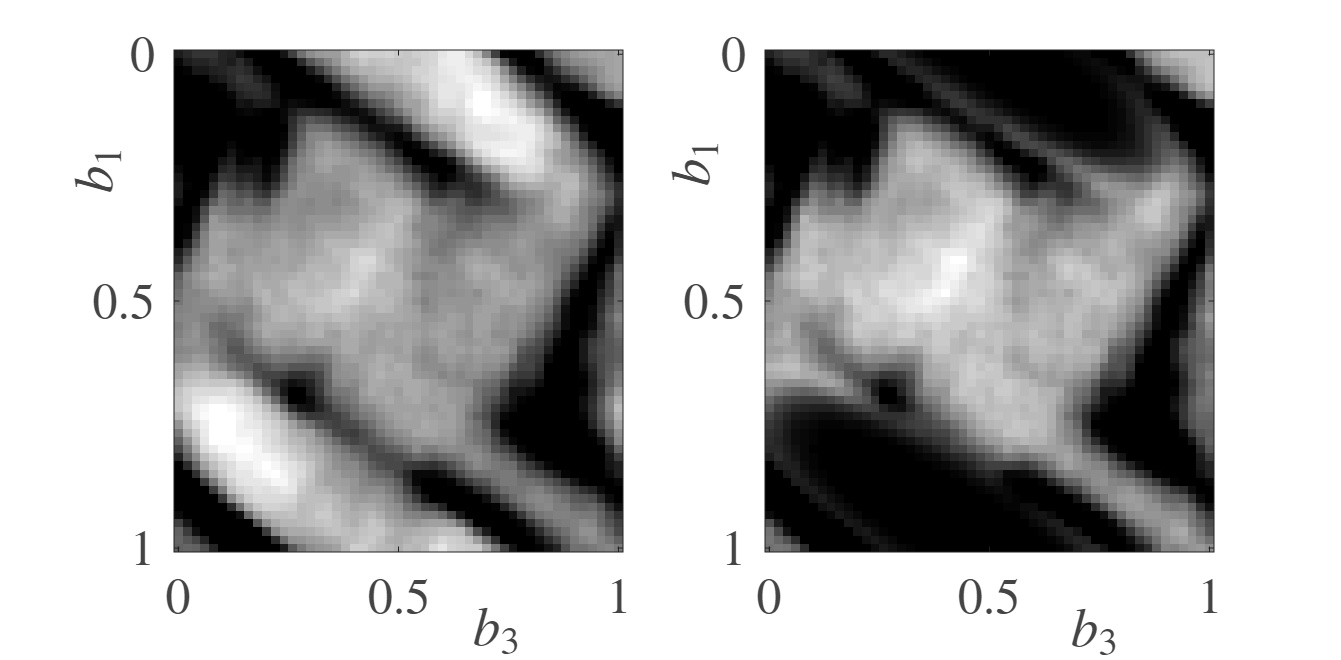}
        \caption{}
    \end{subfigure}
 \end{center}
 \caption{Sections of the raw periodic cell (left column); Same sections after masking the segmented warp yarns (right column)
 }
    \label{fig:erase_warp_period}
\end{figure}

The first step involves removing the warp yarns from the volume by erasing elliptical regions centred on their annotated positions in each slice. These ellipses, shown in \Cref{fig:erase_warp_period}, are defined using fixed cross-sectional dimensions chosen to closely approximate the average shape of the warp yarns, thereby defining forbidden regions where no overlap with weft yarns can occur. Although this operation only provides an approximate representation of the actual yarn cross-sections, its purpose is primarily to avoid weft yarn overlap with predetermined warp yarns and facilitate the identification of the remaining weft regions. The benefit of this masking procedure is illustrated in \Cref{fig:erase_warp_period}, where the remaining weft yarn patterns become significantly more distinguishable after warp removal. It is important to note that the unit cell periodicity is preserved throughout the erasure procedure. When a trajectory centre lies close to a cell boundary, and the corresponding ellipse extends beyond the domain, the ellipse is wrapped across the opposite boundaries to ensure continuity and maximise coverage within the periodic cell.

The second step is to replicate the periodic unit cell twice in the three $(\mathbf{b}_1, \mathbf{b}_2, \mathbf{b}_3)$ directions to more easily render the yarns continuity.  Considering a larger volume is motivated by the difficulty of segmenting yarns near the cell boundaries, where only partial cross-sections are visible.

\begin{figure}[H]
    \centering        \includegraphics[width=0.6\textwidth]{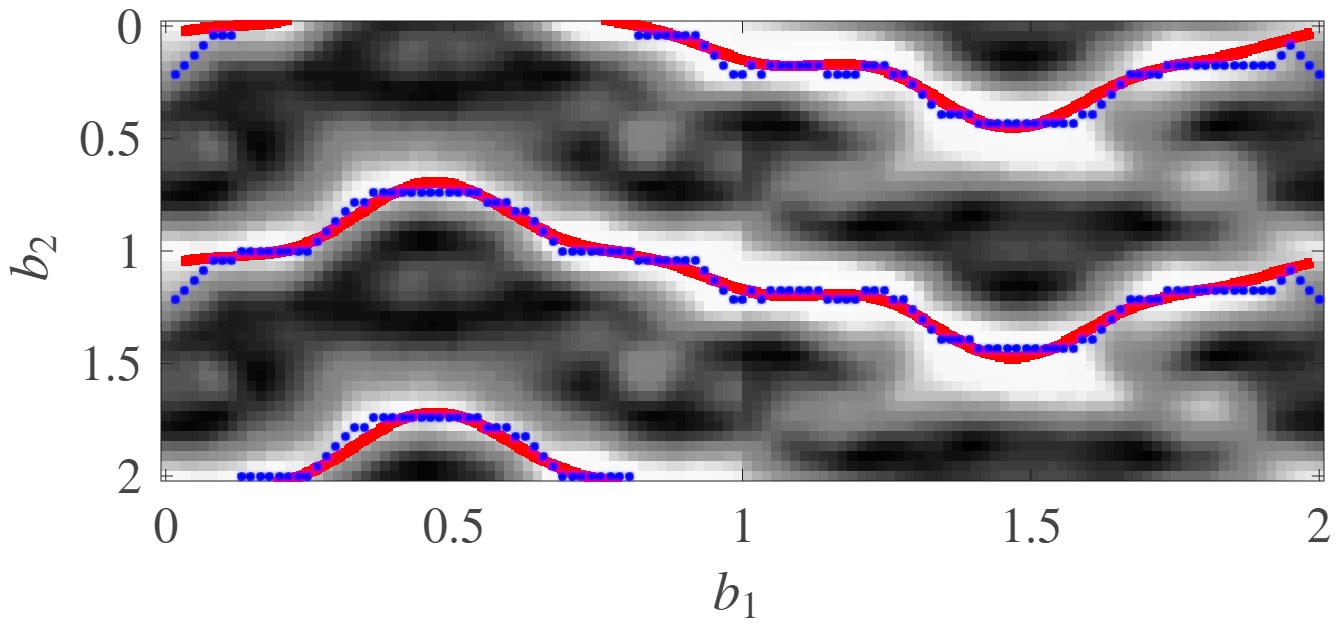}
    \caption{Weft yarn trajectory detection.  Visualisation of the heatmap along with the detected points (blue) and the fitted trajectory (red).}
    \label{fig:periodic_volume_annots_weft}
\end{figure}

\begin{figure}[H]
   \centering
   \includegraphics[width=0.6\textwidth] {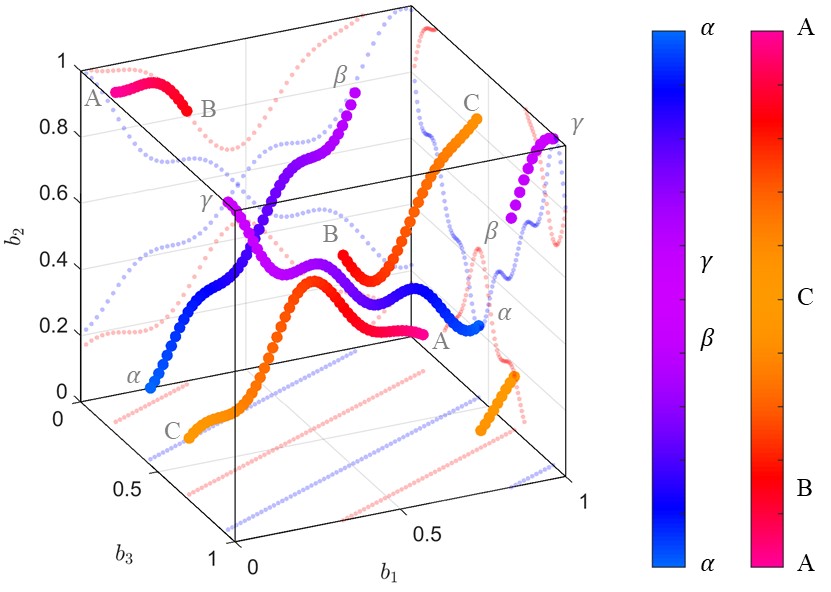}
    \caption{Sinusoidal fit of two tracked weft yarns in the unit cell; their projections onto the planes are shown to illustrate the periodicity.  The A-B-C labels indicate the wrap sections for the first yarn, while the $\alpha$-$\beta$-$\gamma$ do so for the second yarn.}
   \label{fig:annot_curves_weft_combined}
\end{figure}

Weft yarn detection is then performed on the replicated volume.  A cross-correlation between the fundamental mode and the volume is computed, yielding a three-dimensional heatmap.  Viewing this heatmap along the $\mathbf{b}_1$ orientation provides an orthogonal view in which the weft yarns appear as high-intensity regions, corresponding to the local maxima of the heatmap.  Weft yarn trajectories are then extracted by applying Dijkstra's algorithm \cite{Dijkstra_algo} on the corresponding planes, where periodicity is easily implemented. Each resulting path is then fitted with a periodic smooth function (a sine function with the expected periodicity and a few harmonics) using least squares,  yielding a continuous and regular representation of the trajectories.
\Cref{fig:periodic_volume_annots_weft} illustrates the fitted yarn trajectories on a 2D plane, while
\Cref{fig:annot_curves_weft_combined} shows the final yarn paths for the two weft yarns in the unit cell.

\section{Registration with a periodic textile}

\subsection{Theoretical Framework}

Digital Volume Correlation (DVC) is used to register a deformed volume onto a reference one.  In this study, the reference configuration $f(\mathbf{x})$ corresponds to the ideal periodic volume, while the real tomographic image is treated as the deformed configuration $g(\mathbf{x})$.  The registration is primarily accomplished with a transformation $\mathbf{x}\to \mathcal F(\mathbf{x})=\mathbf{x} + \mathbf{u}(\mathbf{x})$, so that $g( \mathcal F(\mathbf{x}))=f(\mathbf{x})$.

The method is originally based on the conservation of grey levels between the two states.  It minimises the $L_2$-norm of the volume difference after registration (the so-called residual) to achieve optimal alignment.
In the present case, a spatially variable affine correction of grey levels is allowed to account for slight tomographic artefacts (such as cupping due to beam hardening) \cite{MENDOZA2019735} and optimised simultaneously with the displacement field.

This goal can be stated as finding the fields $\mathbf{u}(\mathbf{x})$, ${b}(\mathbf{x})$ and ${c}(\mathbf{x})$ that minimise the squared norm of the residual $R(\mathbf{x})\equiv \tilde{f}(\mathbf{x}) - \tilde{g}(\mathbf{x})$
\begin{equation}
    \mathcal{T}=\Vert R(\mathbf{x}) \Vert^2
\end{equation}
with
\be\ba{rcl}
    \tilde{g}(\mathbf{x}) &=& g(\mathbf{x} + \mathbf{u}(\mathbf{x})) \\
    \tilde{f}(\mathbf{x}) &=& f(\mathbf{x}) \cdot (1 + {c}(\mathbf{x})) + {b}(\mathbf{x} )
\ea\label{eq:corrections}
\ee

The fields $(\mathbf{u},b,c)$ are discretised using a finite element formulation, and the resulting non-linear minimisation problem is solved using a Gauss–Newton iterative scheme.  To improve convergence and ensure physically meaningful displacement fields, especially in regions with low texture, a mechanical regularisation based on the equilibrium gap method is applied \cite{MENDOZA201927,Claire2004EquilibriumGap}.
This regularisation introduces elastic constraints that balance registration accuracy and mechanical admissibility through a characteristic regularisation length scale.

\subsection{Implementation and Convergence Strategy}

A multi-scale iterative strategy based on progressive regularisation refinement is used to ensure robust convergence while minimising the registration residual~\cite{MENDOZA2019735}.

The displacement field is first initialised using a rigid body translation estimated from the global motion between the reference and deformed volumes. The optimisation starts with a large regularisation length $l = 1600$~voxels, equal to the depth of the analysed volume, which strongly constrains the displacement field and provides a smooth initial solution. The regularisation length is then progressively reduced through successive stages according to the sequence $l = 1600$, $800$, $600$, $400$, $200$, $100$, and $50$~voxels. At each stage, the displacement field corresponding to the lowest residual is retained and used as the initial guess for the next level. Each correlation stage is carried out until either the norm of the displacement update falls below $0.01$ or a maximum of 100 iterations is reached.

This multi-scale regularisation strategy ensures progressive convergence while maintaining numerical stability, particularly in challenging registration scenarios with limited texture or significant deformation gradients.

The chosen discretisation parameters moreover ensure that the effective kinematic resolution remains smaller than the characteristic textile periodicity in all relevant directions, such that only very local features could potentially be affected by further refinement.

The final displacement field represents an optimal balance between matching accuracy and mechanical plausibility, with the residual minimised across the entire volume.

\begin{figure}[H]
    \centering
    \includegraphics[width=0.8\textwidth]{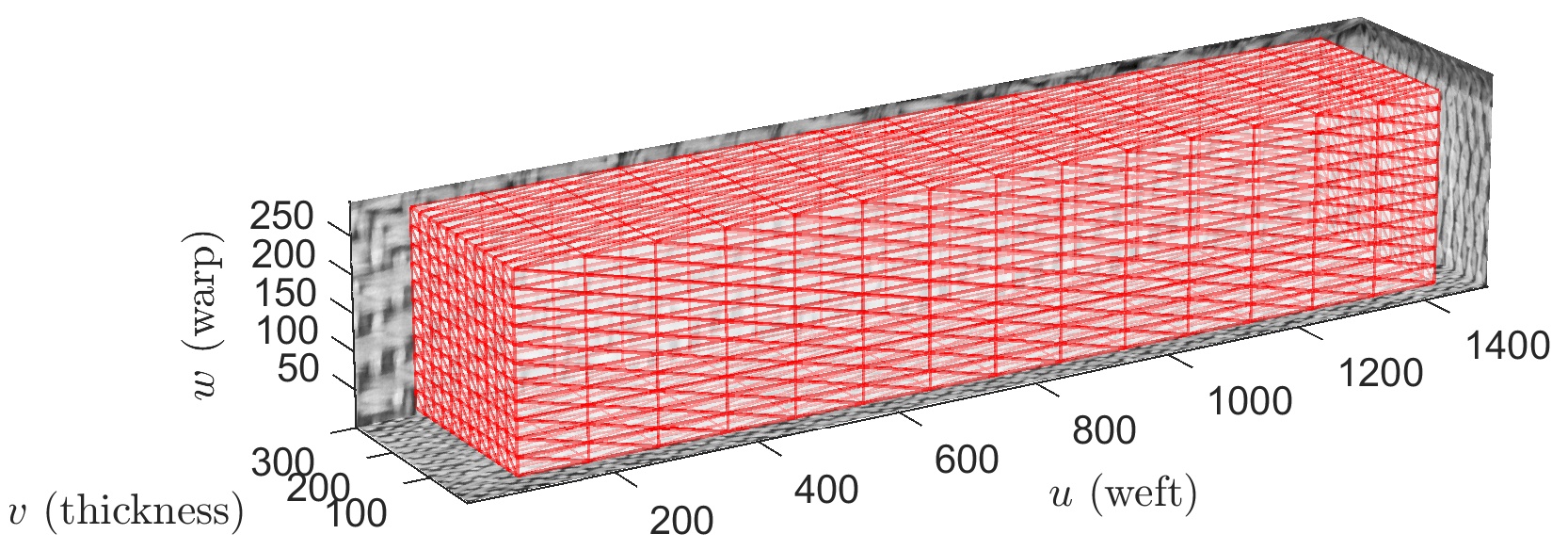}
    \caption{Constructed mesh over the region of interest.}
    \label{fig:mesh_over_volume}
\end{figure}

A structured finite-element mesh was constructed within the 3D image volume to serve as the spatial discretisation for the DVC analysis.
The region of interest was discretised in a {$100 \times 20 \times 25$~voxel} grid composed of tetrahedra elements.
\Cref{fig:mesh_over_volume} shows the obtained mesh.

It may be worth comparing these dimensions to those of the unit cell (see Eq.~\ref{eq:period}).
First, the dimension in the thickness direction (25~voxel) is very close to that of the unit cell (23~voxel).  Next, regarding the in-plane orientation, the unit cell directions do not precisely align with the warp and weft orientations.  Nevertheless, the element dimensions are approximately 60\% larger along the weft direction (100~voxel) and much smaller (about one third of the period) along the warp direction (20~voxel).

While the kinematic description is coarser along the weft direction, the weaving is very regular in this direction.  Moreover, most of the expected strain occurs along the warp direction, where the element size is smaller than that of the unit cell.  Hence, the constructed mesh allows a very fine description of the modulation of the periodic pattern for registration.

\section{Registration results}
\label{sec:results_dvc}

The DVC analysis is performed using the Correli 3.2 library~\cite{leclerc20153}.
The complete MATLAB execution of iterations takes 30 minutes on a standard dual-core CPU.

\begin{figure}[H]
    \centering
    \includegraphics[width=0.5\linewidth]{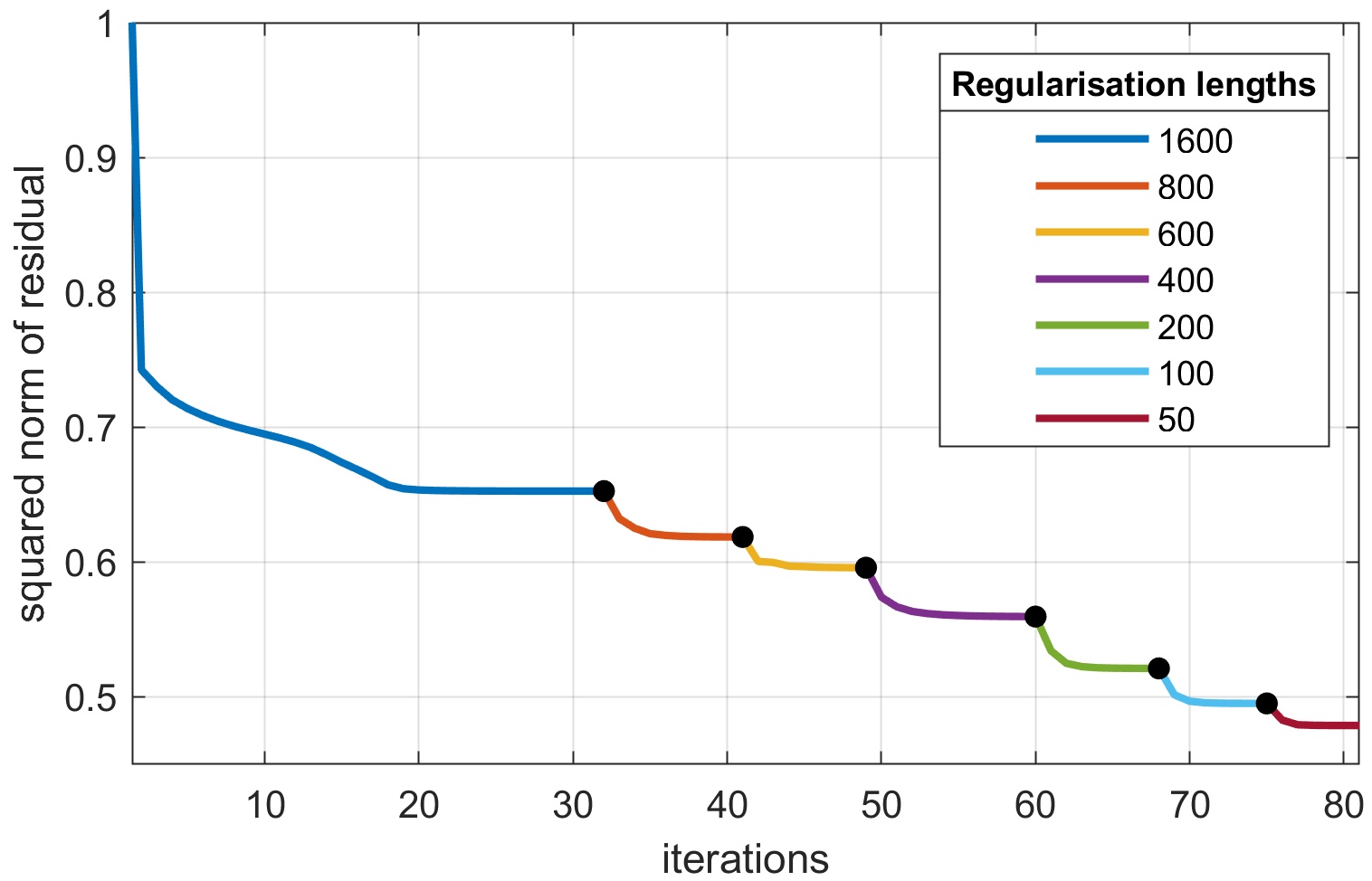}
    \caption{Evolution of residual norm vs. iteration where the different regularisation lengths are indicated by the colour code}
    \label{fig:residual_convergence}
\end{figure}

As shown in~\Cref{fig:residual_convergence}, the squared norm of the residual decreases progressively over iterations and regularisation lengths, reflecting the refinement of the displacement field and the smooth convergence of the procedure.

\begin{figure}[H]
    \centering
    \includegraphics[width=0.8\linewidth]{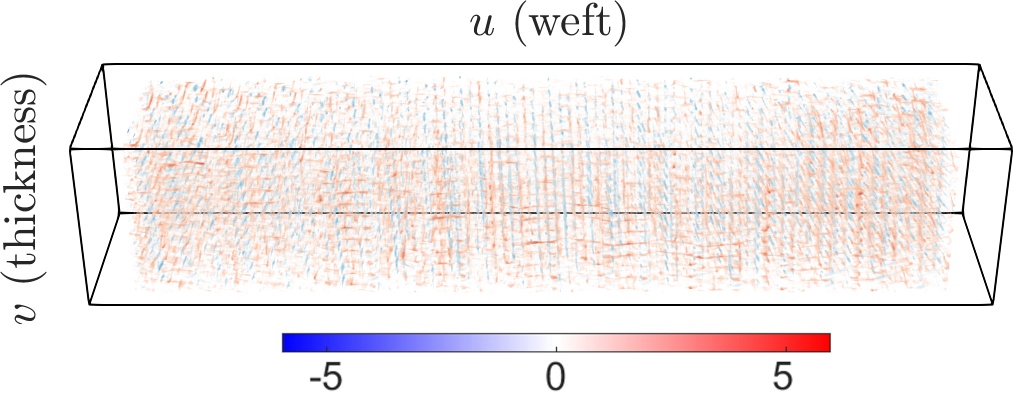}
    \caption{Volume visualisation of the final residual, $R(\mathbf x)$}
    \label{fig:residual_volume}
\end{figure}

One of the merits of DVC is its easy access to quality indicators.  Indeed, after registration, the difference between reference and corrected deformed images, the so-called ``residual field'' or ``residual image'',  clearly indicates where registration was successful and where it may have failed.
This residual is shown in \Cref{fig:residual_volume}. It exhibits small, localised and near-stationary residuals, mainly confined to the boundaries of individual yarns. These discrepancies are expected, as the kinematic description is designed to register the overall textile rather than the fine-scale details of each yarn.  Thus, the final residual field shows no ill-behaviour and confirms that the global textile deformation has been properly captured.  As such, the DVC registration onto the built periodic medium is deemed reliable.

\begin{figure}[H]
    \centering
    \includegraphics[width=.85\textwidth]{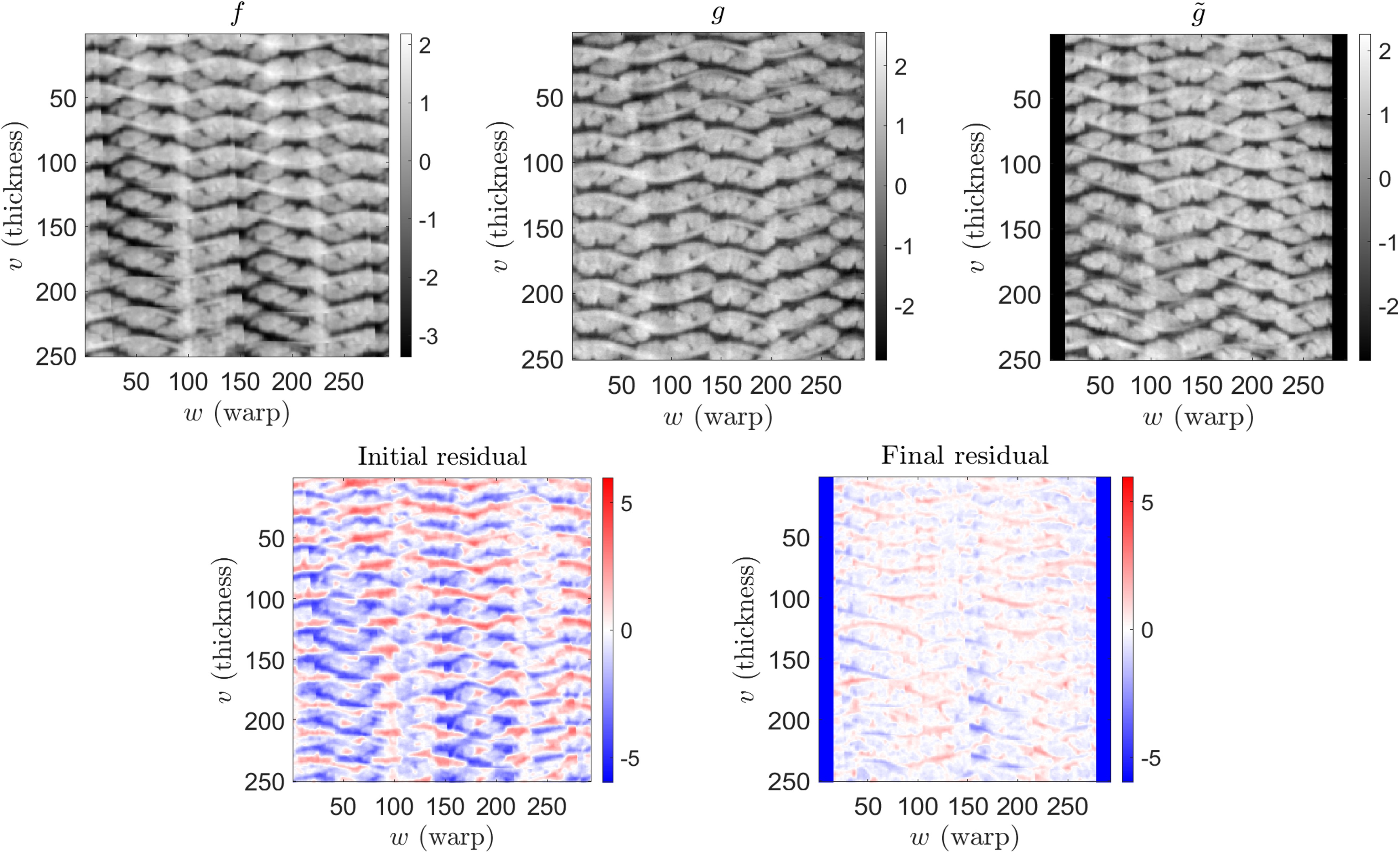}
    \caption{Slice visualisation of the DVC analysis perpendicular to the weft direction.
    Top row: ideal volume $f$ (left), deformed volume $g$ (middle) and corrected deformed volume $\tilde{g}$ obtained by applying the displacement field $\mathbf{u}(\mathbf{x})$.
    Bottom row:  initial residual field $(f - g)$ (left) and residual field $R(\mathbf x)$ (including grey-level corrections) after registration (right). }
    \label{fig:gtilde_residual}
\end{figure}

\begin{figure}[H]
    \centering
    \includegraphics[width=.85\textwidth]{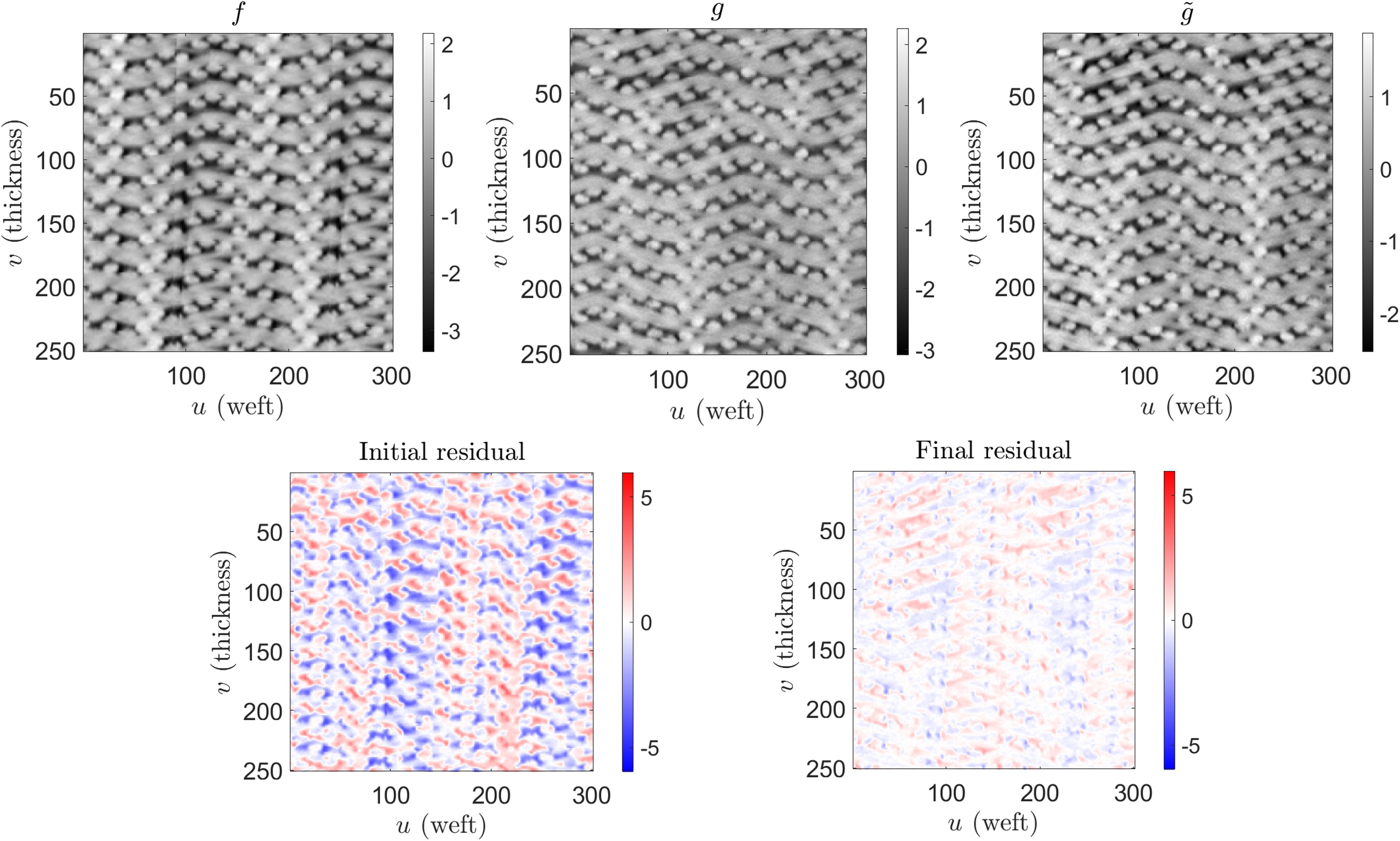}
    \caption{Slice visualisation of the DVC analysis perpendicular to the warp direction.
    }
    \label{fig:gtilde_residual_warp}
\end{figure}

\begin{figure}[H]
    \centering
    \includegraphics[width=.85\textwidth]{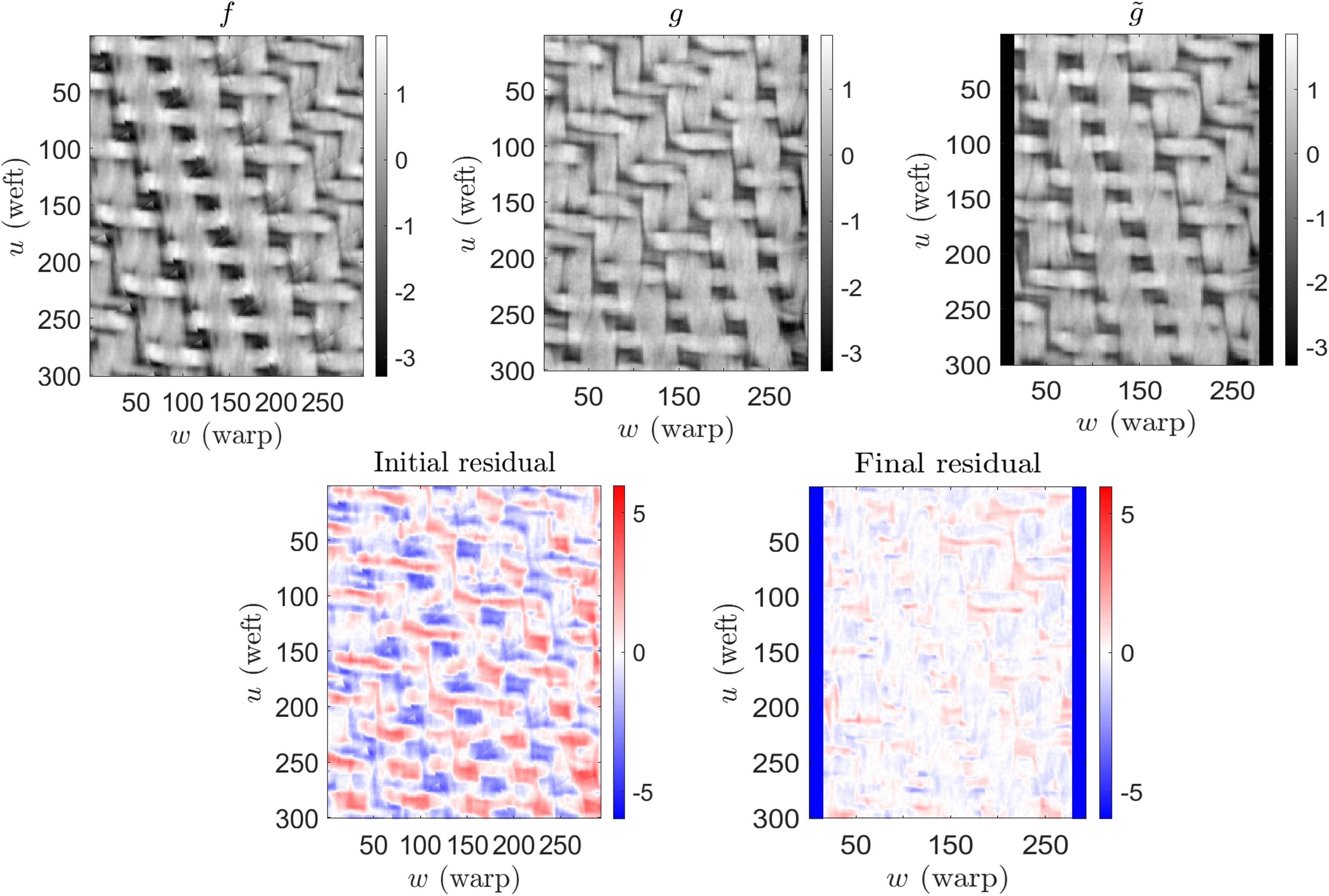}
    \caption{Slice visualisation of the DVC analysis perpendicular to the thickness direction.
    }
    \label{fig:gtilde_residual_thickness}
\end{figure}

Three orthogonal slices of the volumes (periodic reference $f(\mathbf{x})$, real $g(\mathbf{x})$, and corrected real $\tilde g(\mathbf{x})$ images) are shown together with the corresponding residuals (before and after registration) in \Cref{fig:gtilde_residual,fig:gtilde_residual_warp,fig:gtilde_residual_thickness}. The fact that the final residuals are much fainter than the initial difference and almost stationary shows that the registration was successful, and that no systematic bias arises from limitations in the discretisation or in the freedom allowed for the displacement field.

\begin{figure}[H]
    \centering
    \includegraphics[width=0.8\textwidth]{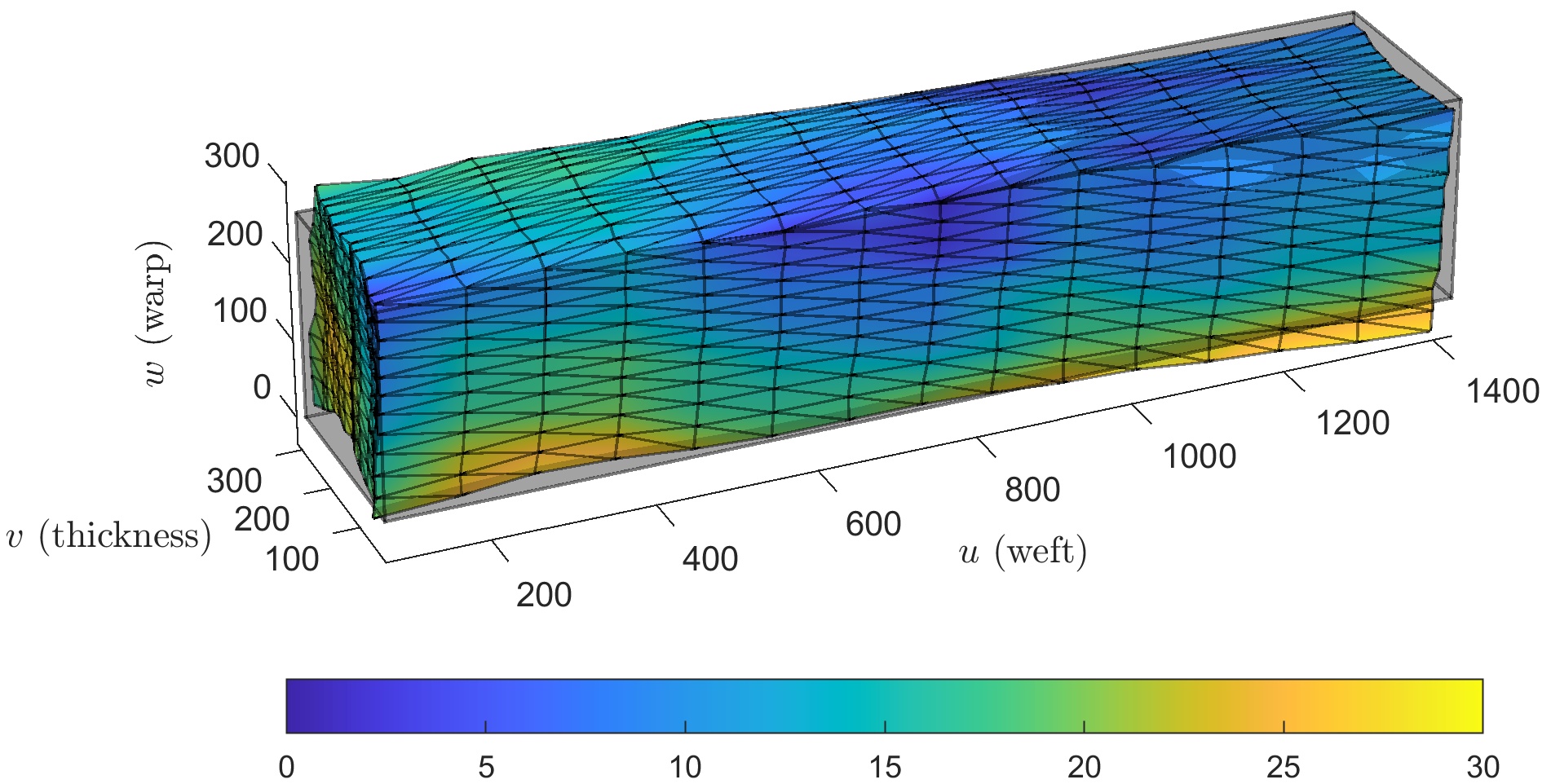}
    \caption{Visualisation of the mesh deformed by the displacement field. The deformation has been magnified by a factor of 2 for visualisation purposes.  The scale indicates the magnitude of displacement in voxels.}
    \label{fig:mesh_deformed}
\end{figure}

\begin{figure}[H]
    \centering
    \includegraphics[width=\linewidth]{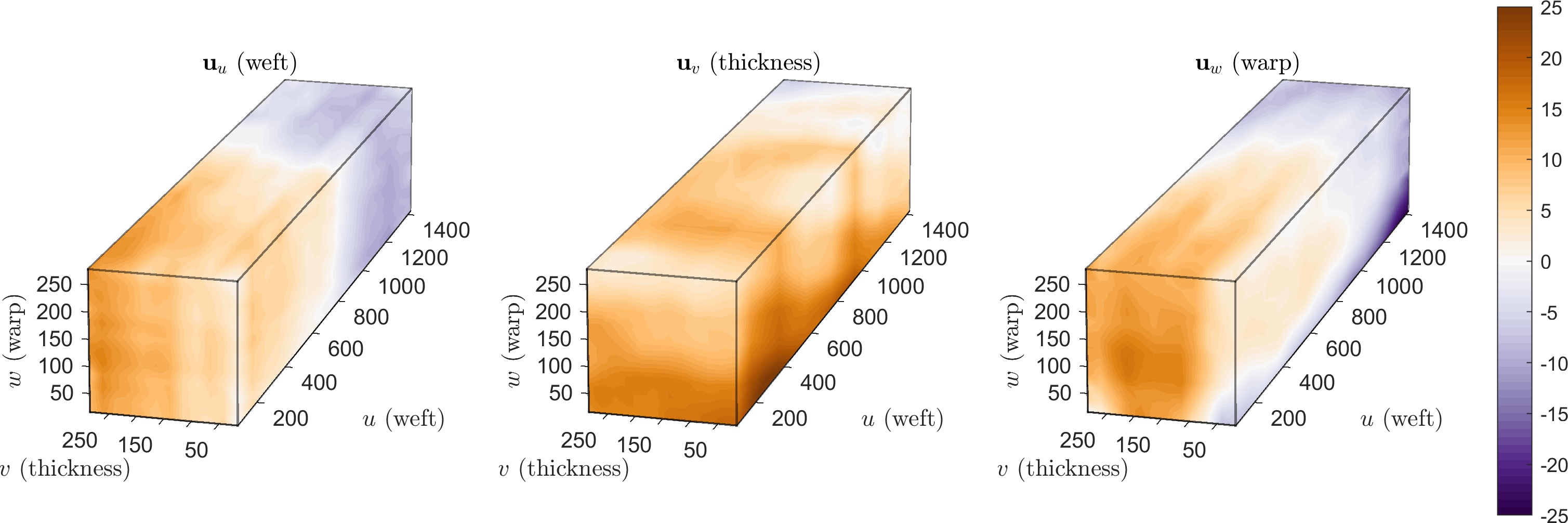}
    \caption{Displacement fields obtained at convergence, scale shown in voxel units.}
    \label{fig:fields_displacement}
\end{figure}

\begin{figure}[H]
    \centering
    \includegraphics[width=\linewidth]{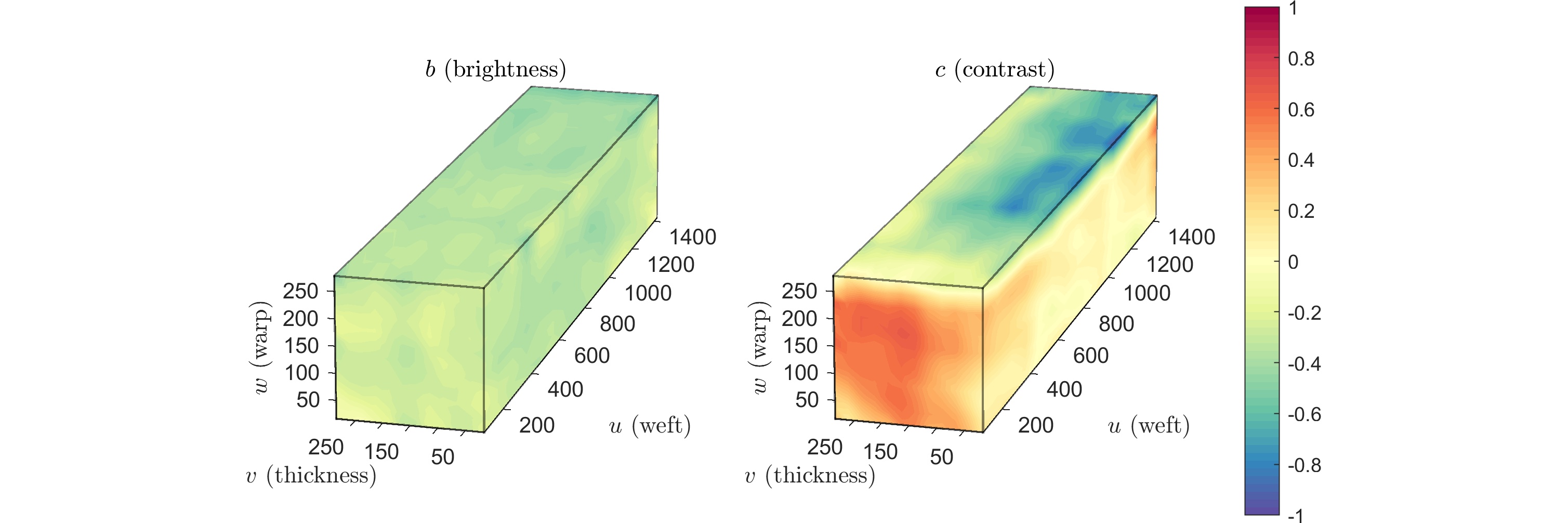}
    \caption{Grey level correction fields obtained at convergence.}
    \label{fig:fields_gray_level}
\end{figure}

Next, \Cref{fig:mesh_deformed} illustrates the obtained displacement field over the deformed mesh, while \Cref{fig:fields_displacement,fig:fields_gray_level} display the different components of the displacement field and the grey-level correction fields.

\begin{figure}[H]
    \centering
    \includegraphics[width=0.4\textwidth]{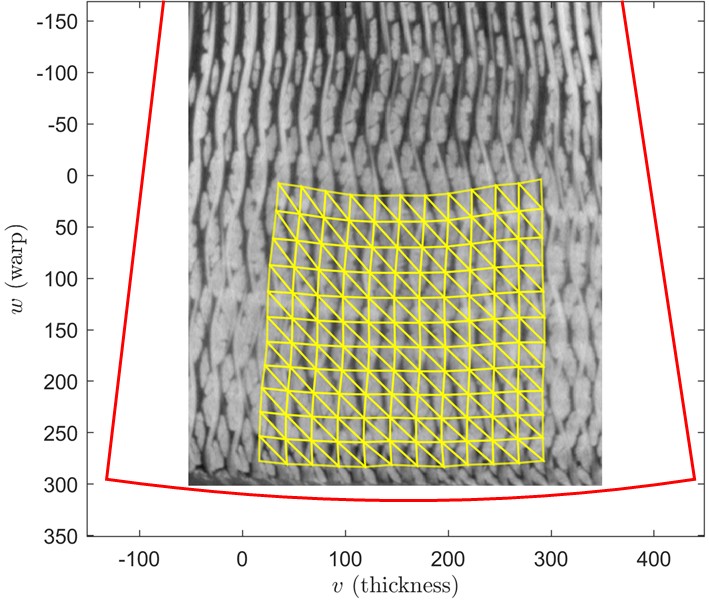}
    \caption{Deformed mesh inside the 3D tomographic volume (normal to the weft direction).}
    \label{fig:weft_mesh_deformed}
\end{figure}

It is worth noting that the largest deformations are localised in the lower region of the volume, corresponding to $w \approx 0$.  This localisation is best observed in \Cref{fig:weft_mesh_deformed}, which shows a mid-plane along the weft direction.  As discussed previously, in the reference periodic volume $f(\mathbf{x})$, the yarns in this region appear as straight columns (by construction), whereas in the real volume $g(\mathbf{x})$ they exhibit a pronounced curvature.  This region ($w \approx 0$) corresponds to the onset of the textile transition from a denser area (dovetail) to a less dense one (blade).  Such a localised modification of the weaving pattern naturally affects the corresponding side of the analysed volume more strongly, while the overall topology remains periodic.

\section{Mapping yarn paths onto the original volume}

The annotated warp and weft trajectories are then propagated throughout the entire periodic volume using the back transformation $\mathbf{T}^{-1}$ to revert to the original $(u,v,w)$ coordinate system.  The result is a fully segmented, periodic volume whose size can be extended without limit.

The inverse transformation $\mathcal{F}^{-1}$, obtained from the Digital Volume Correlation (DVC) between the reference volume $f(\mathbf{x})$ and the real volume $g(\mathbf{x})$ (see \Cref{sec:results_dvc}), can be applied to the tracked periodic trajectories.  This transformation maps the annotations from the ideal periodic configuration to the actual deformed state, correctly accounting for all large-scale geometric variations and distortions present in the real tomographic volume.

\begin{figure}[H]
    \centering
    \includegraphics[width=\textwidth]{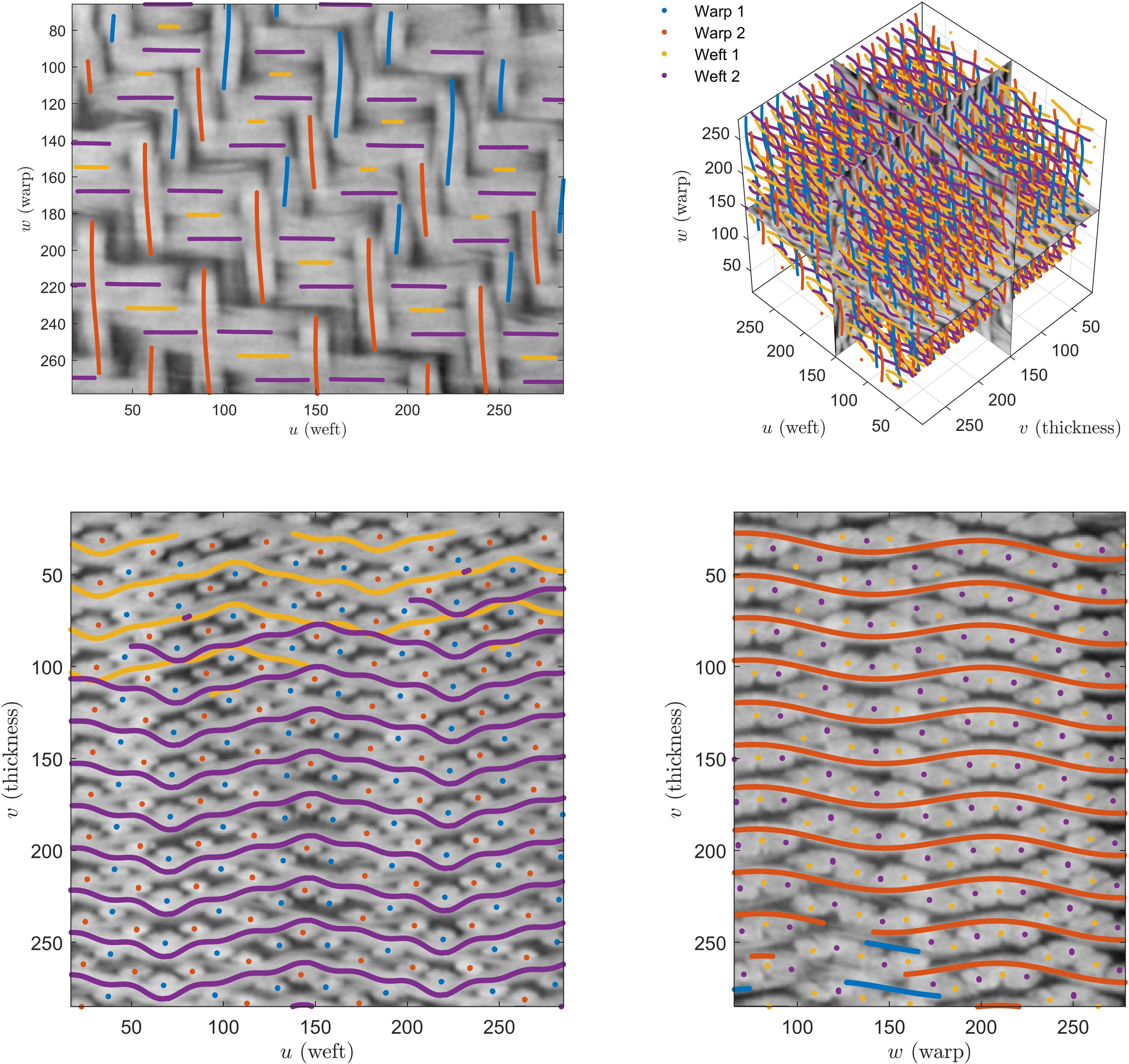}
    \caption{Visualisation of the periodic yarn paths overlaid on the deformed tomographic volume within a small region. The colouring corresponds to the four original yarns in the unit cell.}
    \label{fig:results_final}
\end{figure}

\Cref{fig:results_final} presents a small region of the CT volume, extracted along with the corresponding yarn paths. This figure includes a 3D visualisation as well as the associated mid-planes displaying the intersecting yarn paths. To improve the visibility of the thin yarn lines, the paths are integrated across a few slices around the precise 2D plane. Additionally, as the unit cell contains two warp and two weft yarns, which are then propagated throughout the structure, colours are used to represent these original yarns. This approach results in two distinct sets of warp and weft yarns that are easily visualised, with each set containing multiple yarns that can subsequently be individually labelled in a post-processing step.

\section{Discussion}

Given the lack of annotations for the studied region, direct comparison to a ``ground truth'' is not feasible in the present case.
Consequently, a rigorous quantitative estimation of reconstruction uncertainty or tracking error cannot presently be established over the full analysed volume.
Nevertheless, the validity of the registration-based identification can be robustly assessed by analysing the residual fields.
Indeed, the proposed framework registers an idealised periodic architecture onto a real tomographic volume.
As a result, any incorrect yarn trajectory, erroneous correspondence, or inconsistent geometric mapping would necessarily generate strong and spatially structured residuals after registration, since the local weave pattern would no longer match the observed texture.
In this sense, the residual field provides an intrinsic consistency indicator for the reconstructed architecture.
If the registration procedure were to fail, it would therefore manifest unambiguously as large residual structures correlated with the textile architecture itself.

While the residuals are not strictly zero, the remaining signal mainly corresponds to a slight modulation of the compaction of the periodic unit cell.  Physically, variations in compaction alter the voids between yarns, leading to kinematics that differ from a simple uniform interpolation of grey levels, as currently accounted for.  As a perspective, such alterations of the periodic cell could be explicitly modelled and incorporated into the framework in a manner similar to the existing brightness and contrast ($b(\mathbf{x})$ and $c(\mathbf{x})$ in~\Cref{eq:corrections}) fields.  Apart from these effects, any significant identification error would be clearly revealed in the residual field as ``topological differences''~\cite{MENDOZA2019735} that considerably alter the weave structure.

An additional perspective for quantitative validation would be to apply controlled geometric perturbations to an existing CT volume for which the textile architecture is already known, thereby generating synthetic deformed configurations with an accessible reference solution. Such datasets would enable a more systematic assessment of the identification accuracy and sensitivity of the proposed framework under controlled distortions and imaging conditions. More generally, fully synthetic woven architectures or pseudo-CT data generation approaches, such as those proposed in~\cite{Mendoza_CST_2021}, could also be considered as alternative benchmark strategies.

While a masking strategy similar to that used for warp yarns could, in principle, be extended to weft yarns, the latter are significantly more deformable in the densely packed root region.  As a result, their behaviour cannot be adequately captured by the rigid elliptical model that proves effective for warp yarns, and would require a more elaborate description of their deformation.  Developing such a model constitutes an appealing perspective for future work.
In particular, combining the present periodicity-driven framework with the finer local tracking strategies proposed in~\cite{hafsa2026} could provide a natural extension toward richer descriptions of yarn geometries in complex industrial regions.

Beyond trajectory reconstruction, the proposed framework also induces idealised yarn cross-sections through the propagation process.

While local cross-sectional variations are not represented explicitly, the objective here is primarily the large-scale reconstruction of the textile architecture for subsequent numerical analyses, where preserving the global continuity and consistency of yarn paths is of primary importance. Such periodic trajectories could notably serve as input for geometrical reconstruction frameworks based on deformable yarn representations, such as the hollow shell method~\cite{stig2012spatial}, where idealised yarn paths are subsequently expanded and mechanically adjusted to recover realistic textile compaction and yarn deformations.

The non-orthogonality observed locally in the weave architecture remains limited in practice and is consistent with the shear distortions naturally introduced during draping and manufacturing. Such deviations from an ideal orthogonal arrangement are therefore expected in real industrial components and do not constitute a limitation of the proposed framework. Indeed, the registration procedure is specifically intended to capture these geometric distortions through the mapping between the ideal periodic configuration and the real tomographic volume. Moreover, the non-orthogonal basis obtained from the autocorrelation vectors is used directly to construct the affine transformation defining the periodic reference configuration. The textile is therefore numerically aligned with respect to its own local periodic directions rather than to an artificially imposed orthogonal frame.

\section{Conclusion}

This work proposes a novel segmentation strategy for the blade root, a complex and heretofore unexplored region of woven reinforcements.  Due to the high yarn density and interlacing, individual tracking is extremely challenging when relying solely on 2D images.

The core contribution is the explicit exploitation of intrinsic periodicity of the woven architecture.  Although geometrically distorted by structural constraints, the underlying periodic topology remains well-preserved.  By annotating a minimal representative periodic cell, the proposed method propagates this structural information across the full 3D volume, enabling the reconstruction of thousands of warp and weft yarn trajectories with minimal input.
More generally, the framework is expected to remain applicable to other woven composite systems exhibiting comparable near-periodic organisation.

Unlike previous approaches that avoided this region or relied heavily on statistical modelling, the present framework leverages periodic consistency as a structural prior.  This strategy enables robust yarn tracking even in densely packed areas and provides an efficient approach to generating large, coherent annotation datasets, thereby improving segmentation performance and model scalability.

A distinctive strength of the proposed approach lies not only in its ability to reconstruct large-scale textile architectures but also in its capacity to provide intrinsic validation of tracking quality.
This aspect is often difficult to assess in alternative approaches, such as traditional machine-learning and deep-learning-based approaches.  In addition, the method does not rely on manual annotations or lengthy training, which is a significant advantage, particularly for complex regions such as the fan blade root, where manual labelling is extremely challenging and error-prone.

Finally, the computational cost of the approach remains modest.  Segmenting the unit cell takes less than 1 minute on a standard PC.  The full registration analysis is completed in under 30 minutes.  Such time costs should be viewed in relation to manual segmentation.  This task, while feasible, would require many hours, even for experienced annotators, due to the sheer number of yarns present in the studied region.

\section{Acknowledgments}
Hafsa El Herichi acknowledges the support of a PhD grant N\textsuperscript{o} 2022/1494 from ANRT and Safran.

\end{document}